\documentclass[times,twocolumn,final]{elsarticle}
\usepackage{medima}
\usepackage{framed,multirow}

\usepackage{latexsym}

\usepackage{url}

\usepackage{float}

\usepackage{amsmath}
\usepackage{amssymb}
\usepackage{bm}
\usepackage{diagbox}
\usepackage{algorithmic}
\usepackage[caption=false,font=footnotesize]{subfig}
\usepackage{adjustbox}
\usepackage{lipsum}
\usepackage{footnote}
\usepackage{stfloats}

\usepackage{threeparttable}
\usepackage{multirow}

\usepackage{color}
\definecolor{newcolor}{rgb}{.8,.349,.1}

\journal{Medical Image Analysis}

\begin{document}

\verso{Gang Qu \textit{et~al.}}

\begin{frontmatter}

\title{Integrated Brain Connectivity Analysis with fMRI, DTI, and sMRI Powered by Interpretable Graph Neural Networks }%

%\title{Multimodal Brain Connectivity Analysis via Interpretable Graph Neural Networks with fMRI, DTI, and sMRI}%

% \tnotetext[tnote1]{This is an example for title footnote coding.}

\author[1]{Gang \snm{Qu}}

\author[2]{Ziyu \snm{Zhou}}
\author[5]{Vince D. \snm{Calhoun}}
\author[6]{Aiying \snm{Zhang}\corref{cor1}}
%% Third author's email
\author[1]{Yu-Ping \snm{Wang}\corref{cor1}}

\cortext[cor1]{Corresponding author}
\address[1]{Biomedical Engineering Department, Tulane University, New Orleans, LA 70118, USA}
\address[2]{Computer Science Department, Tulane University, New Orleans, LA 70118, USA}
\address[5]{Tri-Institutional Center for Translational Research in Neuro Imaging and Data Science (TreNDS) - Georgia State, Georgia Tech and Emory, Atlanta, GA 30303, USA.}
\address[6]{School of Data Science, University of Virginia, Charlottesville, VA 22903, USA}

% \received{1 May 2013}
% \finalform{10 May 2013}
% \accepted{13 May 2013}
% \availableonline{15 May 2013}
% \communicated{S. Sarkar}

\begin{abstract}
%%%
Multimodal neuroimaging data modeling has become a widely used approach but confronts considerable challenges due to their heterogeneity, which encompasses variability in data types, scales, and formats across modalities. This variability necessitates the deployment of advanced computational methods to integrate and interpret diverse datasets within a cohesive analytical framework. In our research, we combine functional magnetic resonance imaging (fMRI), diffusion tensor imaging (DTI), and structural MRI (sMRI) for joint analysis. This integration capitalizes on the unique strengths of each modality and their inherent interconnections, aiming for a comprehensive understanding of the brain's connectivity and anatomical characteristics. Utilizing the Glasser atlas for parcellation, we integrate imaging-derived features from multiple modalities—functional connectivity from fMRI, structural connectivity from DTI, and anatomical features from sMRI—within consistent regions. Our approach incorporates a masking strategy to differentially weight neural connections, thereby facilitating an amalgamation of multimodal imaging data. This technique enhances interpretability at the connectivity level, transcending traditional analyses centered on singular regional attributes. The model is applied to the Human Connectome Project's Development study to elucidate the associations between multimodal imaging and cognitive functions throughout youth. The analysis demonstrates improved prediction accuracy and uncovers crucial anatomical features and neural connections, deepening our understanding of brain structure and function. This study not only advances multimodal neuroimaging analytics by offering a novel method for integrative analysis of diverse imaging modalities but also improves the understanding of intricate relationships between brain's structural and functional networks and cognitive development.
%%%%
\end{abstract}

\end{frontmatter}

%\linenumbers

% \section{Note}
% \label{sec1}
% Please use \verb+elsarticle.cls+ for typesetting your paper.
% Additionally load the package \verb+medima.sty+ in the preamble using
% the following command:
% \begin{verbatim}
%   \usepackage{medima}
% \end{verbatim}

% Following commands are defined for this journal which are not in
% \verb+elsarticle.cls+.
% \begin{verbatim}
%   \received{}
%   \finalform{}
%   \accepted{}
%   \availableonline{}
%   \communicated{}
% \end{verbatim}

% Any instructions relavant to the \verb+elsarticle.cls+ are applicable
% here as well. See the online instruction available on:
% \makeatletter
% \if@twocolumn
% \begin{verbatim}
%  http://support.stmdocs.com/wiki/
%  index.php?title=Elsarticle.cls
% \end{verbatim}
% \else
% \begin{verbatim}
%  http://support.stmdocs.com/wiki/index.php?title=Elsarticle.cls
% \end{verbatim}
% \fi

% \subsection{Entering text}
% \textcolor{newcolor}{\bf There is no page limit.}

\section{Introduction}
\label{sec:introduction}  % \label{} allows reference to this section
Advancements in multimodal neuroimaging have revolutionized our understanding of the human brain by providing a harmonized view of its structural and functional information \citep{yan2022deep}. This comprehensive approach enables simultaneous analysis of the brain's anatomy, connectivity, and activity, deepening our understanding of brain function and cognition by capturing a wider range of brain activity and interactions. Additionally, such integrative investigations are vital for exploring the intricacies of learning, memory, language, and emotional regulation, and are instrumental in identifying patterns and biomarkers  \citep{qu2021ensemble, wang2024multiview, liu2024multi} and deviations across developmental stages that relate to cognitive processes  \citep{uludaug2014general, qu2023interpretable, sui2011discriminating}. At the heart of multimodal neuroimaging are functional magnetic resonance imaging (fMRI) \citep{glover2011overview,  wang2021functional, wang2024deep}, diffusion tensor imaging (DTI)  \citep{o2011introduction}, and structural magnetic resonance imaging (sMRI)  \citep{symms2004review}. By combining these modalities, researchers leverage the strengths and mitigate the weaknesses inherent to each modality  \citep{sui2014function}. For instance, fMRI provides insights into brain activity and functional networks  \citep{orlichenko2022latent} by mapping regions active during cognitive tasks. However, its reliance on hemodynamic responses as proxies, combined with limited temporal resolution, restricts its efficacy in capturing instantaneous neuronal dynamics and providing insights into the physical pathways of the brain. Structural connectivity (SC) from DTI  \citep{finger2016modeling} maps the brain's stable anatomical networks but can be compromised by the complex organization of fibers and susceptibility to imaging artifacts. In contrast, sMRI yields detailed morphological insights  \citep{rykhlevskaia2008combining}. However, its capacity to uncover the dynamic interactions of functional brain networks remains limited. In addition, the exploration of the biological mechanisms underpinnings that mediate the interconnections between SC and functional dynamics is understudied. This examination can elucidate the fundamental biological mechanisms by which the anatomical structures of the brain support or constrain its functional manifestations. For instance, studies have demonstrated that regions with high SC often exhibit synchronous functional activities, suggesting a clear "structure determines function" relationship between the physical connections of neurons and their collective functional outputs \citep{honey2009predicting}. Furthermore, disturbances in structural pathways correlate with altered FC, influencing the pathophysiology of diverse neurological disorders and disabilities\citep{piantoni2013disrupted, shu2016disrupted, schaechter2023disruptions}. These findings highlight the importance of investigating structure-function coupling through multimodal data to uncover the neuroscientific and biological mechanisms governing the interactions between structural connectivity and functional networks, and their impact on cognitive functions.

Our study aims to integrate fMRI, sMRI, and DTI for simultaneous examination of the brain, which presents substantial methodological challenges.
The integration of these modalities is complicated by the high-dimensional nature of neuroimaging data, disparate spatial and temporal resolutions, and data heterogeneity—the variability in data types, scales, and formats. This complexity requires sophisticated methods to preserve the intricate topology of neural networks and ensure that the combined modalities accurately reflect both the structural and functional aspects of the brain. Recent literature has underscored the superiority of integrating multimodal neuroimaging data over the utilization of single modality data in the detection of pathological brain anomalies  \citep{sui2013three, zhu2014fusing, stampfli2008combining} and the prediction of phenotypes  \citep{qu2021ensemble} by leveraging the complementary strengths of various imaging modalities  \citep{xiao2022distance, wang2024multiview}. For instance, Zhuang et al.  \citep{zhuang2019multimodal} investigate extracting unique features from each modality to build predictive models. However, this strategy may not fully encapsulate the complex, interrelated dynamics \citep{xu2025deepspatiotemporalarchitecturedynamic} and synergies that exist between the modalities, potentially limiting the comprehensiveness of the predictive analysis. Moreover, there has been a shift towards adopting purely data-driven methodologies  \citep{zhu2022multimodal, shi2020graph, hu2021interpretable, qu2021brain, patel2024explainable}, incorporating advanced computational models to enhance predictive performance. While these approaches have shown promise in terms of accuracy, they frequently overlook the incorporation of established neuroimaging knowledge  \citep{wang2023dynamic, zhou2024interpretable}, thus treating neuroimaging data comparably to natural images without recognizing the unique characteristics and requirements of neuroscientific data analysis. This oversight could lead to the underutilization of critical neuroscientific principles that could otherwise inform and refine the modeling process. A significant academic discourse also revolves around the challenge of model interpretation within this context \citep{hofmann2022towards, orlichenko2022phenotype, chen2024explainable}. Many contemporary models engage in the extraction of high-level features, which, due to their complexity, become opaque and challenging for human interpretation. Even when post-hoc interpretative techniques are applied to elucidate the workings of these models, the resulting explanations often deviate significantly from neuroscientifically relevant insights. This divergence underscores a critical gap in aligning machine learning interpretability with meaningful neuroscientific inquiry, highlighting the need for methodological advancements that bridge this divide.

To address those challenges, we employ a masked Graph Neural Networks (MaskGNN) framework designed to amalgamate SC, FC, and anatomical statistics (AS) using a unified anatomical atlas  \citep{glasser2016multi}. This approach aims to standardize heterogeneous data to a common scale and structure it within a universal graph, facilitating a comprehensive analysis across different dimensions of brain connectivity and morphology. These graphs are subsequently integrated through a masked graph neural network  \citep{qu2021ensemble}, which generates a weighted mask to quantify the significance of each edge in the graph, effectively measuring the comprehensive connectivity strength among brain regions. Our methodology stands out in its adaptability across diverse multimodal datasets, employing a flexible strategy for parcellating and integrating data, thereby consolidating diverse connectivity measures into a a consolidated schema. This approach enables profound insights into both functional and structural connectivities, ensuring the preservation of network topology for rigorous brain analysis and providing intrinsic interpretability. Our model is validated on the Human Connectome Project in Development (HCP-D) dataset  \citep{somerville2018lifespan} to cognitive score prediction task. The findings reveal that our model outperforms established benchmarks, indicating a notable advancement in the domain of multimodal brain network analysis. Our model is then employed to discern critical brain connections and anatomical brain regions, elucidating which morphological features are essential for human cognition. These results are not only corroborated by prior research but also yield new discoveries, reaffirming the advantages of our approach. Our primary contributions lie in the interpretability of an integrated graph deep learning framework that combines fMRI, sMRI, and DTI. Notably, we have:
    (a) a versatile framework that fuses data from fMRI, sMRI, and DTI into coherent graphs, enabling simultaneous analysis of functional, structural, and anatomical metrics;
    (b) a comprehensive approach to interpretability. Specifically, we propose a novel adaptation of a previously introduced weighted-mask approach for processing multimodal data, thereby enhancing the model’s ability to identify significant neural connections;
    (c) new insight into the relationship between brain measurements and adolescent cognitive development validated on the HCP-D dataset. The proposed model not only outperforms existing benchmarks but also yields crucial insights on connectivity previously unattainable with single-modality analyses.
    In summary, we illustrate our approach’s versatility in a novel context, thereby underscoring the practical value of the proposed models in brain development study.

\begin{figure*}[htbp!]
    \centering
    \includegraphics[width=0.85\textwidth]{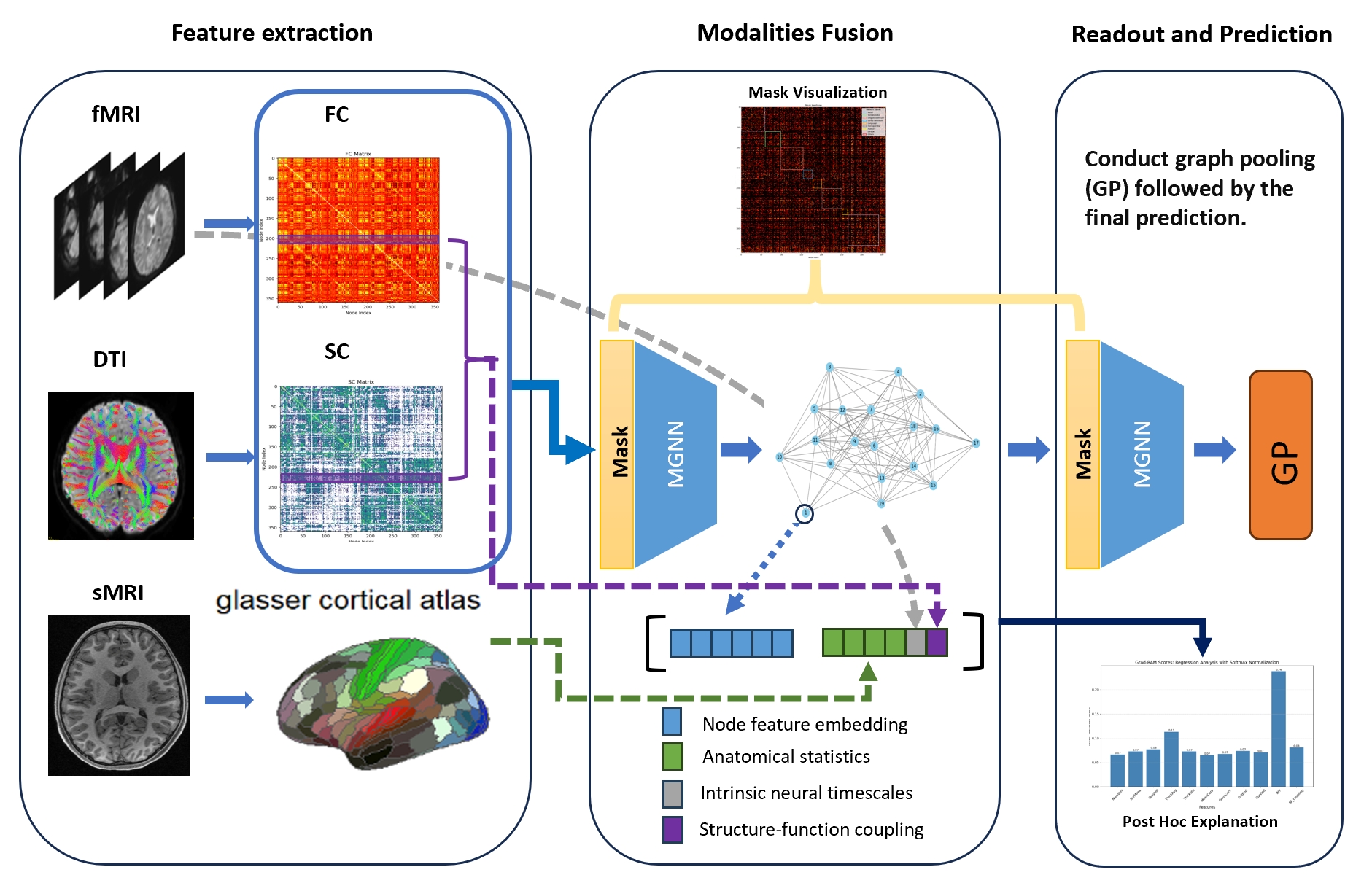}
    \caption{The depiction of the proposed framework: Functional connectivity (FC) and structural connectivity (SC) obtained from fMRI and DTI, respectively, are amalgamated at the nodal level and subsequently fed into the MaskGNN for predictive analysis. In the latent space, embeddings of nodal features are integrated with anatomical statistics (AS) from sMRI, alongside a computation of structure-functional coupling using the FC and SC matrices. The aggregated features are then subjected to MaskGNN embedding, graph pooling, and readout processes. After post-training, the visualization of the uniform mask across MaskGNN layers is achieved, and a post-hoc approach is used to elucidate the contribution of AS.}
    \label{framework}
\end{figure*}

\section{Material and Methods}
\label{sec:method}
\subsection{The Human Connectome Project-Development (HCPD) dataset}

\begin{figure}[htbp!]
    \centering
    \includegraphics[width=0.45\textwidth]{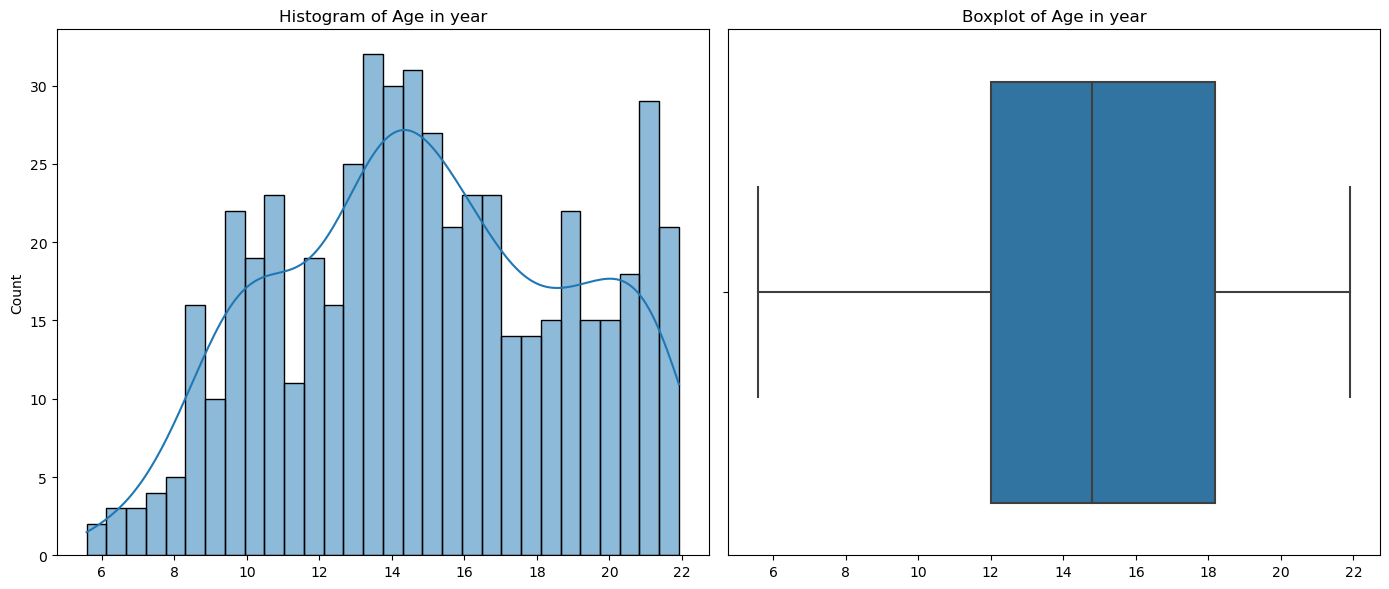}
    \caption{The age distribution of selected subjects.}
    \label{fig:agedis}
\end{figure}

The Human Connectome Project-Development (HCP-D)  \citep{somerville2018lifespan} constitutes a groundbreaking effort dedicated to delineating the progressive maturation of the connectome within a demographically representative cohort of individuals undergoing typical development, spanning ages 5 to 21 years. This study samples a broad geographic, ethnic, and socioeconomic swath of the youth population in the United States, engaging around 650 healthy subjects. A focused subgroup within this cohort undergoes longitudinal observation, especially during the pubertal phase (ages 9 to 17), to rigorously document the patterns of neurodevelopmental changes occurring in this pivotal phase. To ensure consistency and comprehensive coverage, the project adopts a uniform scanning protocol across various locations, utilizing sMRI, DTI, and resting-state fMRI (rs-fMRI). This approach facilitates a comprehensive examination of the brain's structure and function from multiple perspectives. Our study focuses on brain regions excluding the subcortical area and includes subjects with valid data for at least one of three modalities, encompassing a total of 528 subjects. The subject count may vary in the prediction tasks with ablation study due to the possibility of missing modalities. The distribution of age, sex, and race is detailed in Table \ref{tab:concise_distribution} and Fig.\ref{fig:agedis}, respectively.

\begin{table}[ht]
\centering
\caption{Subject Distribution by Sex and Race}
\label{tab:concise_distribution}
\renewcommand\arraystretch{1.5} % This command adjusts the height of table rows for better readability.
\begin{tabular}{|l|c|}
\hline
\textbf{Characteristic} & \textbf{Count} \\ \hline
\textbf{Total Subjects} & 528 \\ \hline
\multicolumn{2}{|l|}{\textbf{Sex}} \\ \hline % Using multicolumn to format the group label
Female & 290 \\
Male & 238 \\ \hline
\multicolumn{2}{|l|}{\textbf{Race}} \\ \hline % Using multicolumn to format the group label
White & 330 \\
Black or African American & 60 \\
More than one race & 86 \\
Asian & 37 \\
Others & 15 \\ \hline
\end{tabular}
\end{table}

\noindent Imaging Preprocessing: We followed the HCP minimal preprocessing pipelines \citep{glasser2013minimal} for s-MRI, DTI, and rs-fMRI. (1) s-MRI: Briefly, structural images are corrected for gradient nonlinearity distortions, intensity inhomogeneity correction. Images were rigidly registered and resampled into alignment with an averaged reference brain in standard space.
(2) rs-fMRI: Preprocessing steps included motion correction, iterative smoothing, motion parameter regression, and rigorous frame censoring \citep{zhang2022decoding} based on framewise displacement (FD) thresholds.
(3) d-MRI: d-MRI preprocessing, implemented in MRtrix \citep{cruces2022micapipe, tournier2019mrtrix3}, including denoising, distortion and motion corrections, co-registration, multiple types of tissue extraction and streamline analysis to facilitate the calculation of SC metrics with the same regions of interest (ROIs).

\subsection{Multi-modal Image-derived Features and Integration}

To mitigate the issue of heterogeneity within multiple neuroimaging modalities, the Glasser atlas \citep{glasser2016multi} is applied to standardize the parcellation of all imaging modalities. Thus, we have a unified graph framework with consistent nodes representing 360 ROIs. This approach facilitates the creation of a neuroanatomical map of the human neocortex, corresponding to multiple imaging modalities. Specifically, the Glasser atlas employs a gradient-based cortical parcellation approach, utilizing an array of multi-modal data, including architectural information from T1w/T2w imaging and cortical thickness maps, task-based and resting-state fMRI, connectivity patterns, and topographical organization. By integrating these diverse data sources, the atlas delineates cortical areas with exceptional precision. Besides, initial areal boundaries are identified based on co-localized gradient ridges across modalities, ensuring a robust, data-driven yet expert-validated mapping process. This semi-automated approach is further refined with machine learning classifiers trained on multi-modal feature maps to automate areal delineation and identification in individuals. Moreover, the methodology prioritizes minimal smoothing and employs multimodal surface matching (MSM) for cortical registration, focusing on areal features over folding patterns to enhance subject alignment without overfitting.

In this study, features reflecting both region-of-interest (ROI)-level and connectivity-level properties of the brain were extracted and analyzed.

\noindent\textbf{Connectivity-level Features:} FC and SC. From rs-fMRI, FC matrix is calculated as the Pearson's correlation between time-series sequences of a pair of ROIs. To generate SC, we first generate a tractography with 10 million streamlines using the iFOD2 algorithm \citep{smith2012anatomically}. Next, spherical deconvolution informed filtering of tractograms (SIFT2) \citep{smith2015sift2} is applied to reconstruct the whole brain streamlines weighted by cross-sectional multipliers. The reconstructed cross-section weighted streamlines are then mapped to the Glasser atlas to form the SC matrix.

\noindent\textbf{ROI-level Features.} 1) Anatomical statistics (AS) obtained from structural MRI (sMRI) are categorized under surface morphology and volumetric  measures, providing quantitative insights into the morphological attributes of brain's cortical structures. These include:
\begin{itemize}
\item Surface Morphology and Volumetric Measures:
\begin{itemize}
    \item Number of Vertices  \citep{kim2016development}: Within neuroimaging, vertices denote the discrete points on the cortical surface, often derived from structural MRI data.
    \item Surface Area  \citep{jha2019environmental, fernandez2016cerebral}: The surface area delineates the total extent of the cortical mantle.
    \item Gray Matter Volume  \citep{gennatas2017age}: Gray matter volume signifies the aggregate volume of neuronal cell bodies and dendrites within the cerebral cortex, elucidating regional differences in neuronal density and synaptic connectivity.
\end{itemize}
\item Metrics Detailing Cortical Thickness (mean and standard deviation)  \citep{dahnke2013cortical}: Metrics detailing cortical thickness encompass measures quantifying the distance between the outer pial surface and the inner boundary surface of the cortex.
\item Curvature  \citep{pienaar2008methodology}:
\begin{itemize}
    \item Mean and Gaussian Curvature: Mean curvature provides a global assessment of cortical surface curvature, while Gaussian curvature quantifies local surface curvature properties, capturing deviations from flatness in orthogonal directions.
    \item Intrinsic Curvature Index: The intrinsic curvature index encapsulates local variations in cortical curvature that are independent of global shape transformations, indicating fine-scale cortical morphology.
\end{itemize}
\item Folding Index  \citep{shimony2016comparison}: The folding index delineates the degree of cortical folding, comparing observed surface area with the theoretical surface area of a smooth cortex, shedding light on cortical morphogenesis
\end{itemize}
\noindent2) Intrinsic neural timescales (INT) \citep{golesorkhi2021brain, watanabe2019atypical, wolff2022intrinsic}: The INT, estimated through the magnitude of autocorrelation of neural signals from rs-fMRI time series, quantifies the duration that neural information is stored in a local circuit. In contrast to FC, the heterogeneity of INT values reflects the fundamental organizational principles of the brain's functional hierarchy, which is broadly relevant to cognitive functions \citep{zhang2024altered}. The calculation of INT is described in Eq.\ref{eq:int}.
\begin{equation}
    INT_{v}= TR \sum_{k=1}^{N_v}\frac{\sum_{t=k+1}^T \left( y_{v}(t) - \overline{y}_{v} \right) \left( y_{v}(t-k) - \overline{y}_{v} \right)}{\sum_{t=1}^T \left( y_{v}(t) - \overline{y}_{v} \right)^2},
    \label{eq:int}
\end{equation}
where $k$ represents the time lag, $T$ denotes the total number of time points, and $y$ refers to the resting-state fMRI signal sequence for each voxel $v$. The voxel-wise INT is then estimated by calculating the area under the curve (AUC) of the autocorrelation function during its initial positive phase. Here, $TR$ is the repetition time, and $N_v$is the lag immediately preceding the first negative value in the autocorrelation function for each voxel $v$ and subject. Following the estimation of voxel-wise INT values, the ROI-specific INT is calculated by averaging these values within each ROI.

\noindent
3) Structure-function coupling \citep{baum2020development}: The structure-function coupling is calculated as the Spearman rank correlation between the SC and FC of each ROI (detailed in the \ref{app:sf}). It measures the spatial correspondence between SC and FC, which describes structural support for functional communication. High coupling occurs when a region's profile of interregional white-matter connectivity predicts the strength of interregional FC. When the Spearman correlation coefficient $\rho$ approaches 1, it signifies a robust positive correlation: SC increases as FC increases.

FC quantifies the temporal correlations between neural activations in different cerebral regions, capturing the dynamic interactions of brain activity. In contrast, SC delineates the anatomical tracts that physically interconnect these regions, providing a static map of neural pathways. AS then offers a detailed examination of the morphological characteristics of these regions, reflecting both their structural integrity and potential functional capabilities.
We apply these analyses consistently across predefined ROIs, which enhances the integration and concatenation of multimodal data at the ROI level. In our study, we combine both ROI and connectivity level features to forge a multidimensional model of brain connectivity. This holistic approach allows us to examine how dynamic functional interactions are underpinned by physical neural pathways and shaped by detailed anatomical features, yielding a deeper insight into the intricate relationships between the brain's structure and function.

FC is normalized using the min-max feature scaling to facilitate comparative analyses across subjects. In contrast, SC is assessed through tractography, which quantifies the number of fiber tracts connecting cortical regions.
The normalization of SC values involves an initial adjustment based on the square root of the product of gray matter volumes in the interconnected regions \cite{hagmann2008mapping}, as detailed in Eq.\ref{eq:sc_normalization} , followed by min-max scaling. To establish graph edges from these connectivities, a threshold value of 0.001 is set, and the 30 largest connections for each node are preserved. Additionally, to address variations in anatomical metrics, a two-step normalization process is applied to AS. This includes a logarithmic transformation to stabilize variance, followed by min-max scaling to ensure all values are confined within a consistent range from 0 to 1.
\begin{equation}
\label{eq:sc_normalization}
\mathrm{SC}_{\text{normalized}}(i,j) = \frac{\mathrm{SC}_{\text{raw}}(i,j)}{\sqrt{V_{\text{GM}}(i) \cdot V_{\text{GM}}(j)}},
\end{equation}
where $\mathrm{SC}_{normalized, raw}$ denotes the normalized and raw SC, $V_{\text{GM}}(i)$ indicates the gray matter volume at the $i_th$ brain region \cite{hagmann2008mapping}.

We combine AS with INT and structure-function coupling using the Spearman rank-order correlation coefficient to create a comprehensive feature vector. For simplicity, we refer this aggregate measure as AS, although it encompasses not only anatomical statistics but also INT and functional-structural coupling. This can capture both static structural details and dynamic functional processes, and quantify the strength and direction of monotonic relationships between structural connectivity and functional activity.

%\subsection{Multimodal Integration}

\subsection{Masked Graph Neural Networks (MaskGNN)}
In our approach, we leverage edge mask learning  \citep{qu2021ensemble} published by us to provide interpretability to our framework on an edge-based level. This methodology diverges from traditional practices where explainability is sought through patterns within individual modalities. Instead, we opt for a unified strategy, retraining our model to simultaneously acquire an edge mask matrix that encompasses subjects across various modalities. This process hinges on the consideration of only undirected graphs, necessitating the edge mask to adhere to symmetry and non-negativity. These constraints are encapsulated in the equation:

\begin{equation}
\bm{\mathcal{M}}=\mathrm{sigmoid}(\bm{V}+\bm{V}^{\mathrm{T}}),
\end{equation}
where $\bm{\mathcal{M}}\in\mathbb{R}^{Q\times Q}$ serves as the mask, with each entry reflecting the important scores attributed to corresponding edges. The matrix $\bm{V}\in\mathbb{R}^{Q\times Q}$ represents the variable we aim to optimize with $Q$ denoting the number of graph nodes. Through the application of the $\mathrm{sigmoid}$ function, we ensure that the elements within $\bm{\mathcal{M}}$ remain positive and are normalized between 0 and 1. Another advantage of mask representation lies in its flexibility in handling large-scale graphs and computational challenges. In such cases (not applicable to our current setting), leveraging low-rank approximations \citep{orlichenko2022latent} and retaining only the upper triangular elements can significantly reduce computational intensity. This innovative approach not only enhances the interpretability of our framework but also ensures a holistic understanding by integrating insights across all considered modalities. Theoretically, our edge mask is applicable across all message-passing graph neural networks, as it adjusts edge weights to tailor neighborhood information aggregation. We employ the Graph Convolutional Neural Network (GCN), as shown in Eq.\ref{mask GCN}, for our specific MaskGNN  backbone module due to its superior performance.

\begin{equation}
    \bm{H}^{l+1}= \mathrm{MaskGNN}(\bm{H}^l)=\phi^l((\bm{\mathcal{M}}+\bm{I})\odot(\tilde{\bm{D}}^{-\frac{1}{2}}\tilde{\bm{A}} \tilde{\bm{D}}^{-\frac{1}{2}})\bm{H}^l\Theta^l),
    \label{mask GCN}
\end{equation}
 where $\mathrm{MaskGNN}$ denotes the forward propagation through the mask GNN layer, while  $\tilde{\bm{A}}=\bm{A}+\bm{I}$ represents the forward propagation of the mask GNN layer and the augmented adjacency matrix with $\bm{I}$  being the identity matrix; $\tilde{\bm{D}}$ denotes the degree matrix corresponding to $\tilde{\bm{A}}$, and $\odot$ signifies the Hadamard product;  $\phi^l$, $\bm{H}^l$ and $\Theta^l$ are the activation function, the feature map and the weight matrix for the $l_{th}$ layer, respectively. Incorporating the identity matrix $\bm{I}$ with the mask matrix $\bm{\mathcal{M}}$ within the GCN architecture ensures an identity mapping, crucial for preventing the graph filter's degeneration into a null matrix when $\bm{\mathcal{M}}$ equals zero. This methodology facilitates controlled information dissemination and tailored neighborhood aggregation, predicated on the learned edge weights; therefore, it preserves the graph's structural integrity and enhances model robustness by maintaining self-connections and mitigating information loss.

The initial layer of the MaskGNN produces a graph embedding $\hat{\bm{H}}^1$, which is then fused with anatomical statistics (AS) $\bm{C}\in\mathbb{R}^{Q\times d_{c}}$, characterized by various morphological measurements with $d_c$ specifying the count of features. This concatenation process occurs at the node level, with each node's feature embedding being combined with the corresponding brain region's AS to guarantee both homogeneity and dimensional compatibility. This fusion is captured by the equation $\bm{H}^1=\hat{\bm{H}}^1\oplus\bm{C}$. Following this fusion, the graph is advanced to the subsequent layer of the MaskGNN and a graph pooling (GP) operation, leading to the final predictive outcome, as shown in Eq.\ref{output}.
\begin{equation}
\hat{\bm{y}} = f(\mathrm{GP}(\mathrm{MaskGNN}(\bm{H}^1))),
    \label{output}
\end{equation}
where $\hat{\bm{y}}$ represents the final predictions, and $\bm{H}^1$ denotes the input feature at the $l_{th}$ layer of the model.

\subsection{Objective function}

To optimize model performance and mitigate oversmoothing, we implement a manifold regularization term to manage the smoothness of node embeddings, represented in Eq.\ref{manifold}:
\begin{equation}
L_{manifold}=\frac{1}{2}\sum_q^Q\sum_{j\in N_q}||\bm{h_q}-\bm{h_j}||_2^2=\mathrm{trace}(\bm{H}^\top\bm{L}\bm{H}),
\label{manifold}
\end{equation}
where $\bm{H}$ represents the node embeddings at the last MaskGNN layer.
The manifold regularization term enforces similarity among embeddings of adjacent nodes, thereby conserving local manifold structures. It quantifies this relationship using the squared Euclidean distance between embeddings of neighboring nodes, fostering continuity and incorporating the graph Laplacian $\bm{L}$ to effectively enforce this smoothness constraint throughout the graph. In addition, the manifold loss is monitored during training with a predefined threshold to prevent overfitting and oversmoothing, ensuring the model’s generalizability while preserving brain network integrity.

In addition to imposing $L_1$ and $L_2$ constraints on the mask $\mathcal{M}$ to promote sparsity, a stringent orthonormality condition Eq.\ref{orthonormality} is enforced. This condition mandates that all rows (and columns) of the mask $\mathcal{M}$ be mutually orthogonal unit vectors, characterized by$||\mathcal{M}_i||=1$ and $mean(\mathcal{M}_i)=0$, thereby ensuring both symmetry and orthogonality within the matrix. Such a constraint significantly enhances the model's ability to learn independent and stable features across different samples, thereby improving generalizability and mitigating the risk of overfitting. Furthermore, the regularization term associated with orthonormality in a symmetric matrix serves to maintain the learned representations close to a set of basis-like, independent features, reinforcing the structural integrity of the model.
\begin{equation}
L_{mask}= \lambda_1 ||\mathcal{M}||_1 + \lambda_2 ||\mathcal{M}||_F^2+\lambda_3||\mathcal{M}\mathcal{M}^\top - \mathbf{I}_Q||_F,
\label{orthonormality}
\end{equation}
where $\mathbf{I}_Q$ is the identity matrix with dimensions matching with those of mask $\mathcal{M}$, and $\lambda_{1-3}$ are the regularization parameters.
Thus, the loss function integral to our proposed architectural is given as follows:
\begin{equation}
L = L_e(\hat{\bm{y}}, \bm{y}) + \alpha L_{manifold} + L_{mask},
\label{final loss}
\end{equation}
where $L_e(\cdot)$ denotes the error in prediction, quantified through cross-entropy in classification scenarios or mean squared error (MSE) in regression task, and $\alpha$ is the regularization parameter in the manifold term. All hyperparameters, including $\lambda_{1-3}$ and $\alpha$, impact both model stability and generalization. We empirically fine-tuned these parameters by conducting a random search across predefined grids and evaluating the results based on the total predictive errors of the validation set.
\subsection{Model Interpretation}
Our framework distinguishes itself through inherent interpretability, achieved by learning masks during model optimization that incorporate multimodal fusion, thereby illuminating the significance of the graph's original connectivity. However, given that mask learning is driven by the downstream predictive task, a smaller degree of sparsity may be expected to ensure optimal predictive performance.  To enhance visualization and feature clarity, we judiciously adjust a visualization threshold, as shown in Eq.\ref{heatmap}.
\begin{equation}
\Tilde{\bm{M}}_{i,j} =
\begin{cases}
\bm{M}_{i,j} & \text{if } \mathrm{sigmoid}(\bm{M}_{i,j}) > \text{threshold}, \\
0 & \text{otherwise},
\end{cases}
\label{heatmap}
\end{equation}
where $\bm{M}\in\mathbb{R}^{Q\times Q}$ represents the learnable mask matrix that is applied to the edges.

In addition to analyzing the graph's connectivity, our interest extends to identifying which anatomical statistics are most pertinent to the predictive task at hand. To achieve this, we employ gradient-based methods, specifically Gradient-weighted Regression Activation Mapping (Grad-RAM)  \citep{qu2021ensemble} and Gradient-weighted Classification Activation Mapping (Grad-CAM)  \citep{chattopadhay2018grad, hu2021interpretable}, to quantify the relevance of each feature that is integrated into the graph embedding within the latent space. These methods facilitate the calculation of the importance score for each feature, providing insights into its respective contribution to the model's predictions.

\begin{equation}
\bm{G}= \frac{\partial\bm{y}}{\partial\bm{H}},
\label{gradients}
\end{equation}
where $\bm{G}\in\mathbb{R}^{Q\times C}$ signifies the gradient matrix with $Q$ being the number of nodes and $C$ the number of features, $\bm{y}$ represents the ground truth, and $\bm{H}$ refers the target features (specifically the $AS$ features in the experiments) used for gradient computation.
Therefore, modulated by the values of AS, Grad-RAM/Grad-CAM is characterized by the interaction between the values and their associated gradients through a product operation.

\begin{equation}
\bm{a} = \frac{1}{Q}\sum_{q=1}^Q\mathrm{ReLU}(\bm{G}_q\odot\bm{H}_q),
\label{RAM}
\end{equation}
where $\bm{a}\in\mathbb{R}^C$ delineates the Grad-RAM/Grad-CAM vector pertinent to AS with the incorporation of the $\mathrm{ReLU}$ function to exclusively preserve those features exerting a positive influence on the ultimate prediction; subscript $q$ indicates the index of a specific node. Following this, normalization of the activation map is executed via the Softmax function. This post-hoc analysis, enhanced through the incorporation of intrinsic masked GNN layer, establishes a robust framework for the interpretation of the model. Notably, it enables the systematic identification of critical brain regions at multiple analytical strata, particularly focusing on the connection (edge) level and the detailed level of individual (node) features for a comprehension of the model's predictive dynamics.
% \begin{table}[]
% \renewcommand\arraystretch{1.0}
%   \begin{center}
%     \caption{Notations and GGT forward propagation.}
%     \label{notations}
% \begin{tabular}{|c|l|}
% \hline
% \textbf{Notation} & \textbf{Description}                                                                   \\ \hline
% \( \bm{H}^{(l)} \)     & Features at the \( l \)-th layer     \\ \hline
% \( \hat{\bm{A}} \)     & Augmented adjacency matrix           \\ \hline
% $\bm{\Theta}$  & Weight matrix
% \\ \hline
% $\bm{AS}$ & anatomical statistics matrix
% \\ \hline
% $\bm{E}$  & GCN-embedded features
% \\ \hline
% \( \hat{\bm{D}} \)     & Degree matrix of \( \hat{\bm{A}} \)  \\ \hline
% \( \bm{W}^{(l)} \)     & Weight matrix for the \( l \)-th layer
% \\ \hline
% \( \sigma \)      & Non-linear activation function
% \\ \hline
% \end{tabular}
% \end{center}
% \end{table}

\section{Experiments}
\label{Experiments}
%\subsection{Human Connectome Project-Development Dataset}

\begin{table*}[htb!]
\renewcommand\arraystretch{2.0}
  \begin{center}

    \caption{Prediction Performance on intelligence scores.}
    \label{Scoreres}
    \begin{adjustbox}{width=\textwidth}
\begin{tabular}{|c|c|c|c|c|c|c|c|c|c|}
\hline
Model        & Modalities     & CCC  RMSE       & P-value          & CCC  MAE        & P-value          & FCC  RMSE       & P-value          & FCC  MAE        & P-Value                   \\ \hline
MaskGNN          & FC             & 17.910 $\pm$  0.118         & \textless{} 0.001 & 14.847 $\pm$ 0.122          & \textless{} 0.001 & 16.382 $\pm$ 0.142         & \textless{} 0.001 & 12.973 $\pm$ 0.107          & \textless{} 0.001         \\ \hline
MaskGNN          & SC             & 19.557 $\pm$ 0.195         & \textless{} 0.001 & 15.305 $\pm$ 0.090          & \textless{} 0.001 & 16.957 $\pm$ 0.0.021          & \textless{} 0.001 & 13.468 $\pm$ 0.045         & \textless{} 0.001          \\ \hline
MaskGNN & FC+SC & 17.580 $\pm$ 0.060 & \textless{} 0.001   & 14.687 $\pm$ 0.059 & \textless{} 0.001   & 16.164 $\pm$ 0.009 & \textless{} 0.001   & 12.989 $\pm$ 0.039 & \textless{} 0.001 \\ \hline
\textbf{MaskGNN}          & FC+SC+AS       & \textbf{14.968 $\pm$ 0.819}          & -                & \textbf{12.095 $\pm$ 0.534}          & -                & \textbf{14.338 $\pm$ 0.754}          & -                & \textbf{11.516 $\pm$ 0.542}    & -                         \\ \hline
GCN          & FC+SC+AS       & 15.654 $\pm$ 0.127          & 0.026               & 12.366 $\pm$ 0.074          & 0.196               & 16.853 $\pm$ 0.110          & \textless{} 0.001               & 13.727  $\pm$  0.096          & \textless{} 0.001                        \\ \hline
GAT          & FC+SC+AS       & 16.230 $\pm$ 0.517         & 0.003  & 12.209 $\pm$ 0.099        & 0.574 & 17.531 $\pm$ 0.307         & \textless{} 0.001 & 13.987 $\pm$ 0.190          & \textless{} 0.001          \\ \hline
GIN          & FC+SC+AS       & 16.978 $\pm$ 1.004         & \textless{} 0.001            & 13.768 $\pm$ 0.924         & \textless{} 0.001            & 17.777 $\pm$ 0.712         & \textless{} 0.001 & 14.907 $\pm$ 0.786         & \textless{} 0.001          \\ \hline
Linear       & FC+SC+AS       & 18.061 $\pm$ 0.047         & \textless{} 0.001 & 15.335 $\pm$ 1.776         & \textless{} 0.001 & 17.092 $\pm$ 0.040          & \textless{} 0.001 & 13.802 $\pm$ 1.776          & \textless{} 0.001          \\ \hline
MLP          & FC+SC+AS       & 17.804 $\pm$ 0.576          & \textless{} 0.001 & 14.473 $\pm$ 0.879         & \textless{} 0.001 & 17.305 $\pm$ 0.520          & \textless{} 0.001 & 14.430 $\pm$ 0.903          & \textless{} 0.001          \\ \hline
\end{tabular}
 \end{adjustbox}
\end{center}
\end{table*}

\subsection{Experimental Setup}
\label{Experimental setup}
\begin{figure}[ht!]
    \centering
    \includegraphics[width=0.45\textwidth]{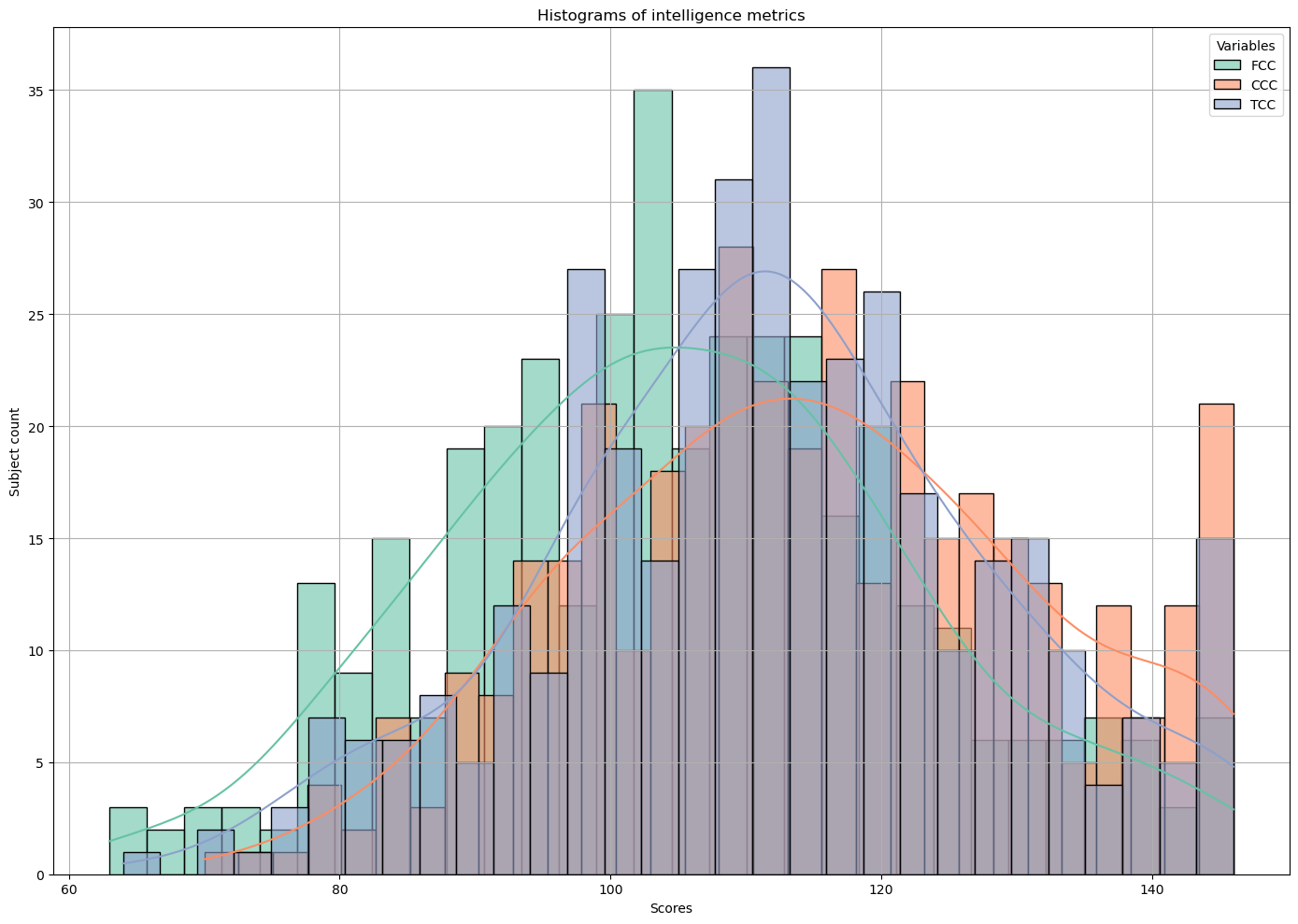}
    \caption{The distribution of intelligence metrics.}
    \label{fig:IQ_his}
\end{figure}

HCP-D encompasses a range of phenotypic measurements, among which intelligence metrics such as fluid intelligence, crystallized intelligence, and total intelligence—a composite measure of the first two—are selected as the supervisory labels for our model. Fluid intelligence is characterized by the capacity to think logically and solve new problems, independent of previously acquired knowledge. It is crucial for adapting to new situations and tackling novel challenges. In contrast, crystallized intelligence involves the application of accumulated knowledge and experience to solve problems. The model is first applied to estimate age-adjusted Crystal Cognition Composite (CCC) and age-adjusted Fluid Cognition Composite (FCC) scores using multimodal neuroimaging data for the prediction task. In the classification task, participants are next categorized into two groups based on extreme age-adjusted Total Cognition Composite Score (TCC) levels: below borderline ($<80$) and very superior ($>130$), highlighting significant differences. These measurements are adjusted for age variations to ensure accurate and reliable comparisons, illustrating the dynamic interplay between the capacity for innovative problem-solving and the utilization of learned knowledge. The distribution of these intelligence metrics is depicted in Fig.\ref{fig:IQ_his}.
The dataset is partitioned into training, validation, and testing subsets at the ratio of $70\%$, $10\%$, and $20\%$, respectively. We construct the graph based on biologically meaningful connectivity patterns, using FC for multimodal tasks and SC for SC-only experiments, selecting the top 30 connections per ROI \citep{qu2020graph} according to the connectivity matrix values. Given the extreme sparsity of SC, a full set of 30 connections for each ROI is sometimes unavailable, in which case we retain the maximum number of existing connections. This thresholding ensures a balance between preserving meaningful structural relationships and computational efficiency. Additionally,  the incorporation of an identity matrix in the mask, as shown in Eq.\ref{mask GCN}, ensures the preservation of self-loops, allowing each ROI to maintain its intrinsic features. Even when certain connections are not explicitly preserved, message passing allows information propagation through k-hop neighbors, ensuring effective feature aggregation and robust graph representation. The model undergoes training on the training set and hyperparameter tuning on the validation set. For the regression task, evaluation metrics, specifically the root mean square error (RMSE) and mean absolute error (MAE), are derived by comparing the predicted and actual test scores within the testing set.  Importantly, despite variations in test scores, all participants are considered healthy, with no physical or cognitive impairments. Bootstrap analysis is employed to evaluate and benchmark the performance of models, aiming to reduce sampling bias with 10 iterations of experiments. Each deep learning model is designed with an initial two-layer structure, leading to a dense readout layer for making predictions. Hyperparameters are optimized on a model-specific basis, employing L2 regularization and drop out to mitigate overfitting across the board. This approach is augmented by an adaptive learning rate, utilizing a ReduceLROnPlateau scheduler with a patience parameter of 10, to dynamically modify the learning rate in response to performance metrics during training and validation phases. For the MaskGNN model, the initial learning rate is established at 0.005, with training parameters set to a batch size of 32 and a maximum of 50 epochs. The L2 regularization coefficient is carefully adjusted to 1e-6 to reduce overfitting, and a sparsity parameter of 30 is used to retain only the largest K neighbor nodes in graph construction, optimizing both model complexity and computational efficiency. Hyperparameter tuning utilizes a random search approach, concentrating on variables including the initial learning rate, batch size, length of training, and the degree of regularization, etc. Moreover, the model integrates distinct regularization terms ($L_1$ and $L_2$) for mask sparsity.

\subsection{Results}
\label{Results}
\subsubsection{Comparative Analysis of Model Predictive Efficacy}
The predictive performance of our model for intelligence metrics is compared against established benchmarks, such as Linear Regression (LR) and Multilayer Perceptron (MLP). Since our framework is general applied to other graph-based deep learning architectures, we also compare different backbone modules such as GCN, Graph Isomorphism Network (GIN) \citep{xu2018how, patel2024explainable}, and Graph attention network (GAT) \citep{vel2018graph, cai2022graph}, with results delineated in Table \ref{Scoreres}. From the table, the MaskGNN model emerges as the paramount model when employing a tripartite combination of modalities - FC, SC, and AS. This superiority is quantitatively supported by achieving the lowest Root Mean Square Error (RMSE) and Mean Absolute Error (MAE) across both CCC and FCC, underscored by significance tests (p-values) derived from t-tests comparing the performance of our MaskGNN model against competing models across repeated experiments, all below a specified threshold.
Moreover, we engage in the classification task to differentiate between groups defined by high and low TCC, as illustrated in Table \ref{classperformance}. This effort substantiates the enhanced predictive capability of our MaskGNN model, which exhibits superior accuracy and Area Under the Curve (AUC).
\begin{figure*}[hbtp!]
\centering
\subfloat[]{
  \includegraphics[width=0.8\textwidth]{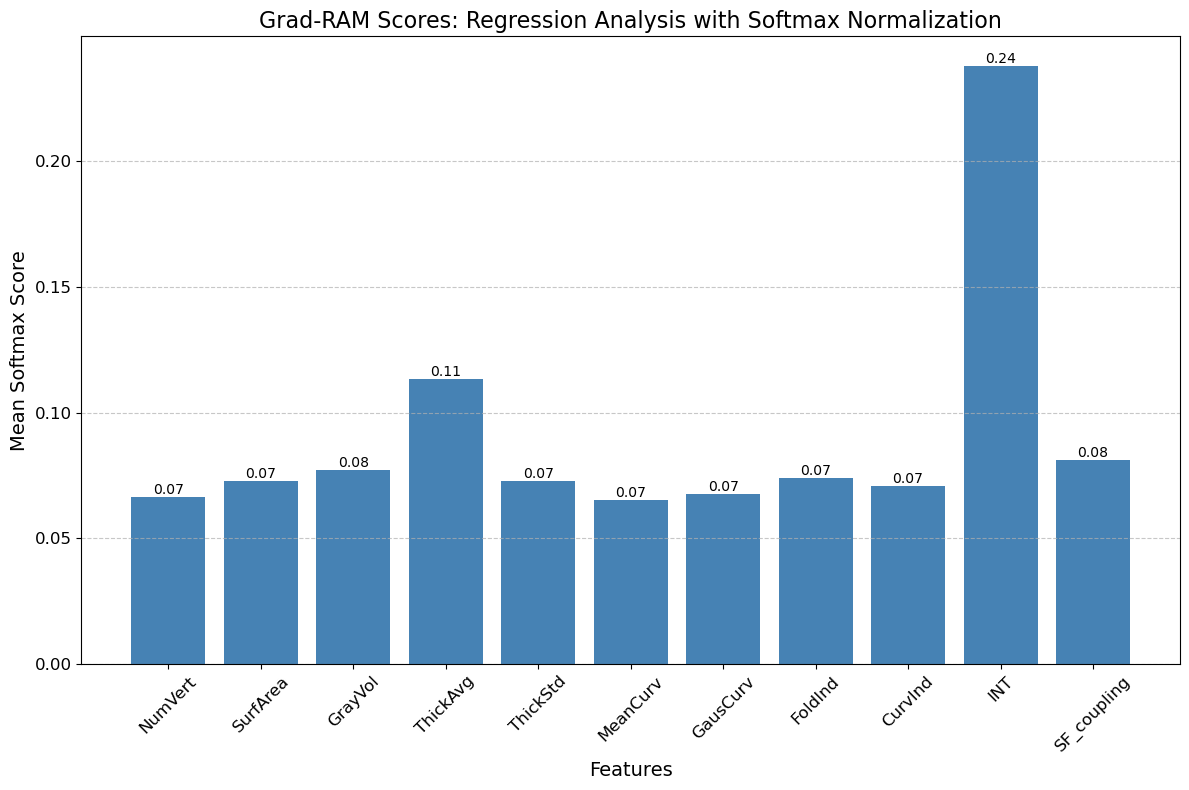}
  \label{fig:gradcam_reg}
}

\subfloat[]{
  \includegraphics[width=0.8\textwidth]{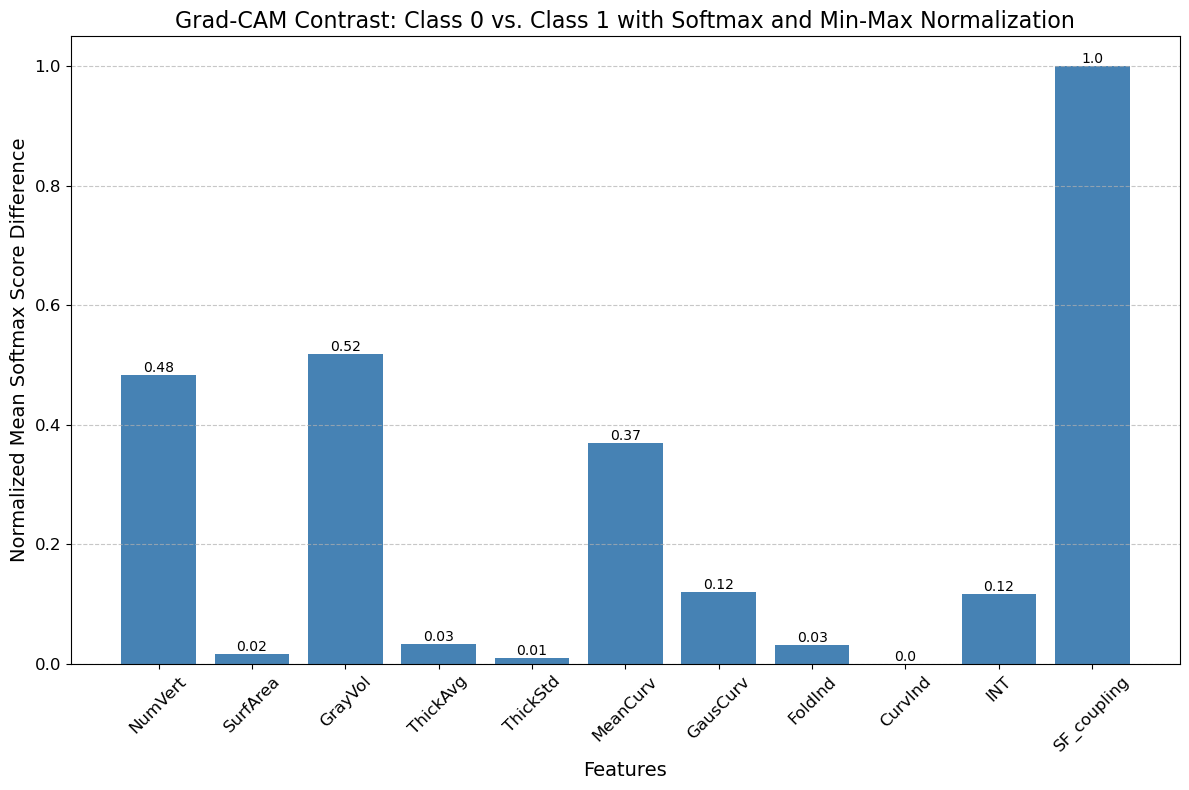}
  \label{fig:gradcam_class}
}
\caption{The use of Grad-CAM and Grad-RAM scores for model explainability: (a) Grad-RAM scores for simultaneous prediction of CCC and FCC; (b) Discrimination of groups using Grad-CAM scores across distinct TCC levels}
\label{fig:gradscore}
\end{figure*}
The analysis highlights performance ranking, showing that MaskGNN outperforms both standard machine learning models and frameworks incorporating different graph-based modules, despite adopting all three multimodal strategies, while achieving lower RMSE and MAE. These findings accentuate the MaskGNN model's capacity for nuanced intelligence score prediction through optimal multimodal data integration.
\begin{table*}[htb!]
\renewcommand\arraystretch{2.0}
  \begin{center}
    \caption{The Performance of Group Classification Based on Intelligence Scores.}
    \label{classperformance}
\begin{tabular}{|c|c|c|c|c|c|c|}
\hline
Model   & Accuracy & P-value & F1-score & P-value & AUC & P-value \\ \hline
\textbf{MaskGNN} &  \textbf{ 0.870 $\pm$ 0.060}       &  -       &   \textbf{0.924 $\pm$ 0.035}         &    -     &    \textbf{0.768 $\pm$ 0.168}        &     -  \\ \hline
GAT     &     0.830 $\pm$ 0.064    &    0.24     &  0.906 $\pm$ 0.038   &    0.36   &      0.624 $\pm$ 0.056       &  \textless{} 0.05      \\ \hline
GCN     &    0.825 $\pm$ 0.033    &    0.09     &    0.903 $\pm$ 0.020     &   0.18       &     0.519 $\pm$ 0.069        & \textless{} 0.05        \\ \hline
GIN     &    0.780 $\pm$ 0.046      &    \textless{} 0.05     &   0.871 $\pm$ 0.029         &   \textless{} 0.05      &   0.646 $\pm$ 0.085          &    0.09     \\ \hline
MLP  &    0.790 $\pm$ 0.030   &    \textless{} 0.05      & 0.882 $\pm$ 0.019          &    \textless{} 0.05      &     0.543 $\pm$ 0.100     & \textless{} 0.05        \\ \hline
Linear     &   0.795 $\pm$ 0.027       &   \textless{} 0.05      & 0.886 $\pm$  0.017       &    \textless{} 0.05     &   0.636 $\pm$ 0.069        &     \textless{} 0.05    \\ \hline
\end{tabular}
\end{center}
\end{table*}
\subsubsection{Ablation Study}
We conduct experiments to evaluate the predictive accuracy of the MaskGNN model using only FC, or SC, and a combination of FC and SC without AS. The outcomes demonstrate a significant difference, as confirmed by the t-test. The ablation study on integrating modalities underscores the pivotal role of synthesizing FC, SC, and AS modalities for enhanced predictive accuracy, as shown in Table \ref{Scoreres}.
Results demonstrate the advantages of multimodal integration that significantly boost predictive performance. The amalgamation of diverse neural data streams—FC, SC, and AS—provides a comprehensive view of the brain's cognitive framework, thereby refining the precision of cognitive intelligence predictions.

\begin{figure*}[htbp!]
    \centering
    \includegraphics[width=0.8\textwidth]{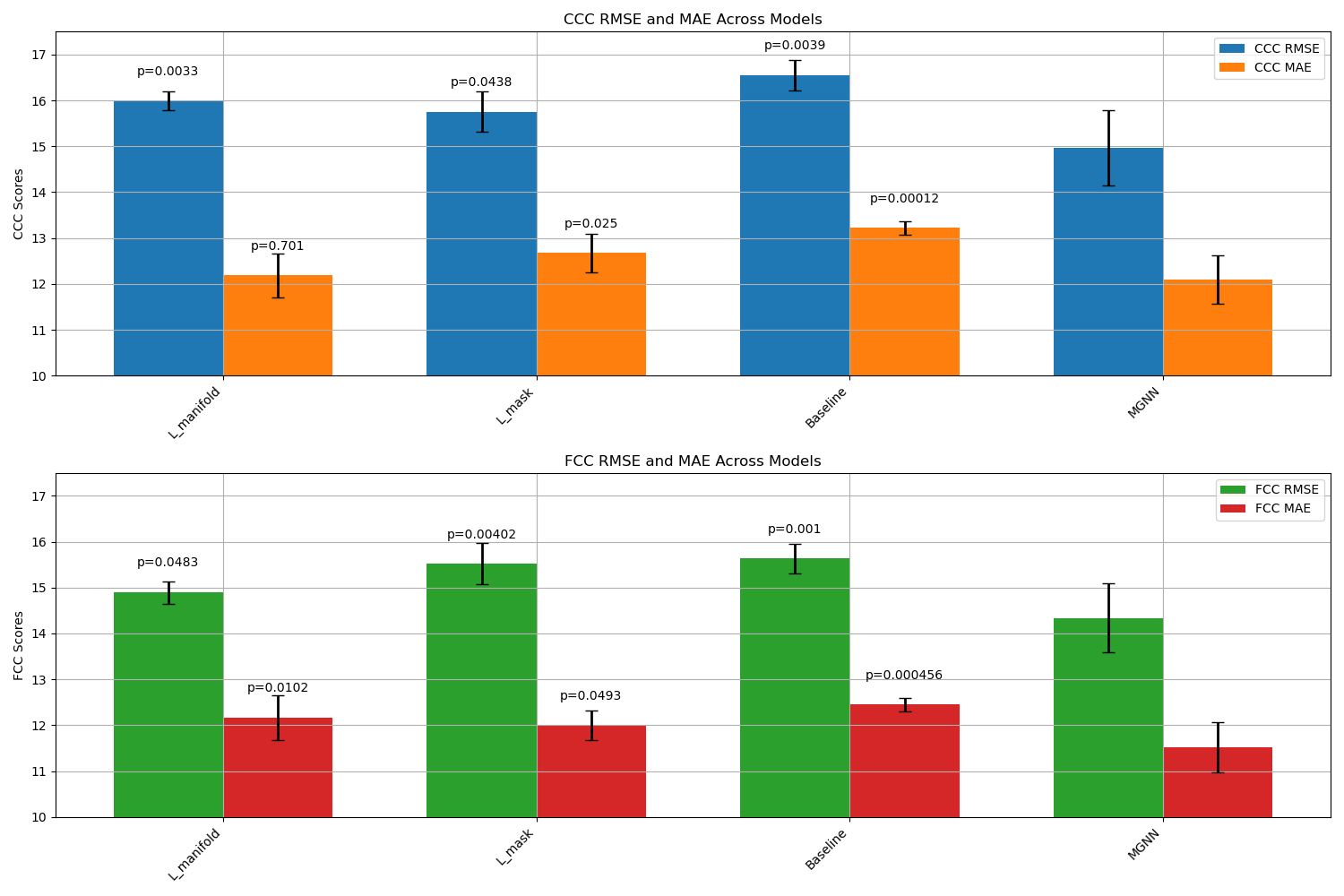}
    \caption{A comparative analysis of predictive performance showing the individual and combined effects of the manifold regularization term ($L_{manifold}$) and mask penalty ($L_{mask}$) on the proposed model, with a baseline scenario for reference. All comparisons are supported by pair-wise t-tests against MGNN, with p-values displayed above each bar except for MGNN, emphasizing significant differences.}
    \label{ablation}
\end{figure*}

Furthermore, we assess the influence of the manifold regularization term and the mask penalty on predictive performance by conducting a comparative analysis across four scenarios: employing solely $L_{manifold}$, solely $L_{mask}$, neither (establishing a baseline), and the fully proposed model. The results of these comparisons, shown in Fig.\ref{ablation}, indicate that all conditions differ significantly in their mean performance compared to the model with both terms included, except for the MAE metric when $L_{manifold}$ is removed ($p=0.7$). However, in that scenario, the RMSE is still significantly different ($p=0.0033$). Overall, both terms substantially affect the performance, but $L_{manifold}$ exerts a more pronounced impact because it directly smooths the embeddings.

\begin{figure*}[hbtp!]
\centering
\subfloat[]{
  \includegraphics[width=0.4\textwidth]{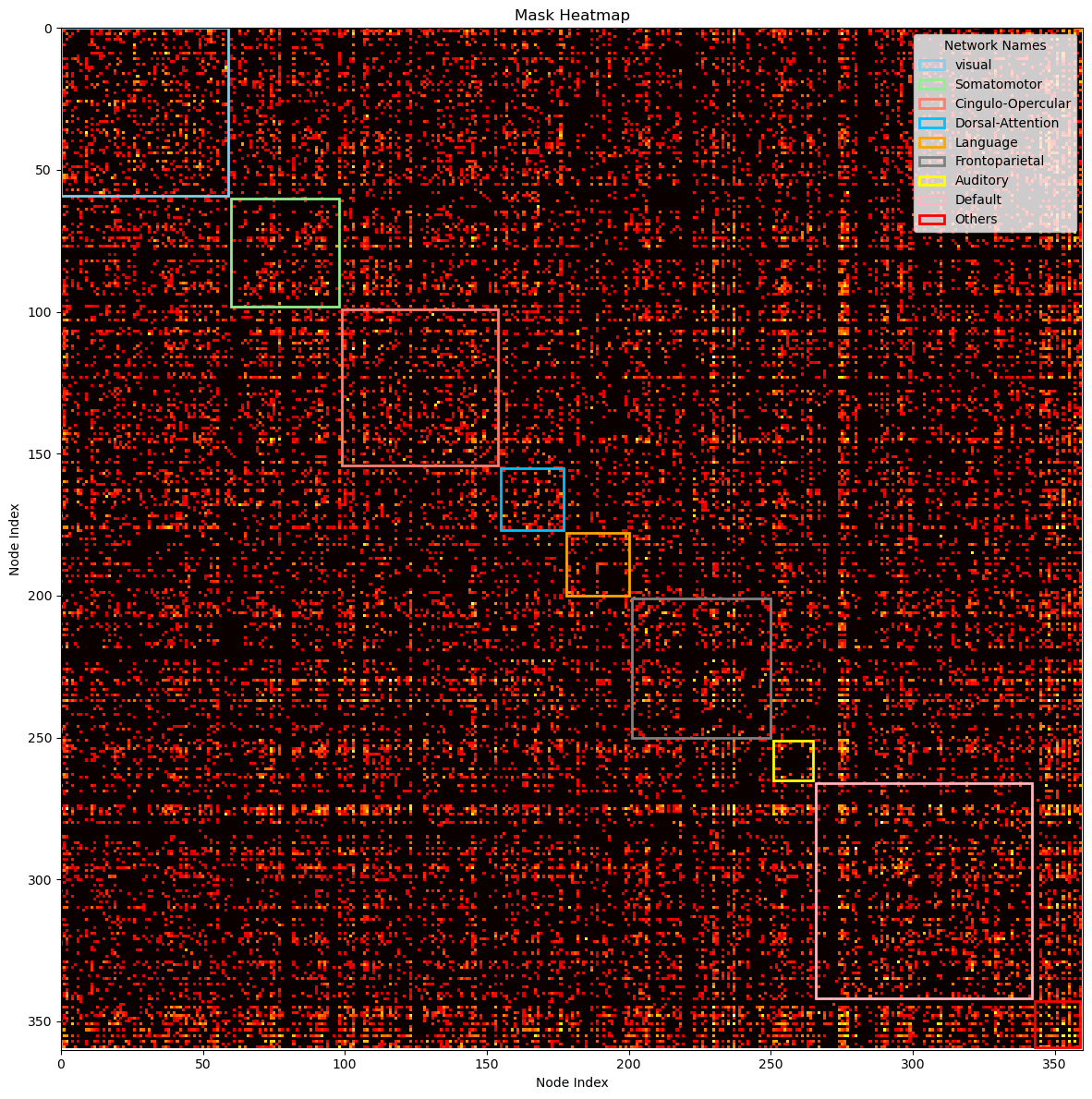}
  \label{fig:mask_reg}
}
\subfloat[]{
  \includegraphics[width=0.4\textwidth]{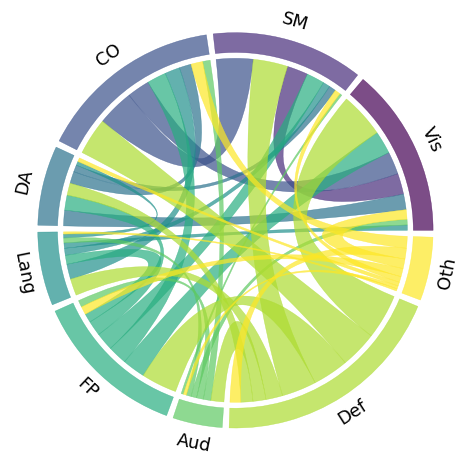}
  \label{fig:chord_diag_class}
}

\subfloat[]{
  \includegraphics[width=0.4\textwidth]{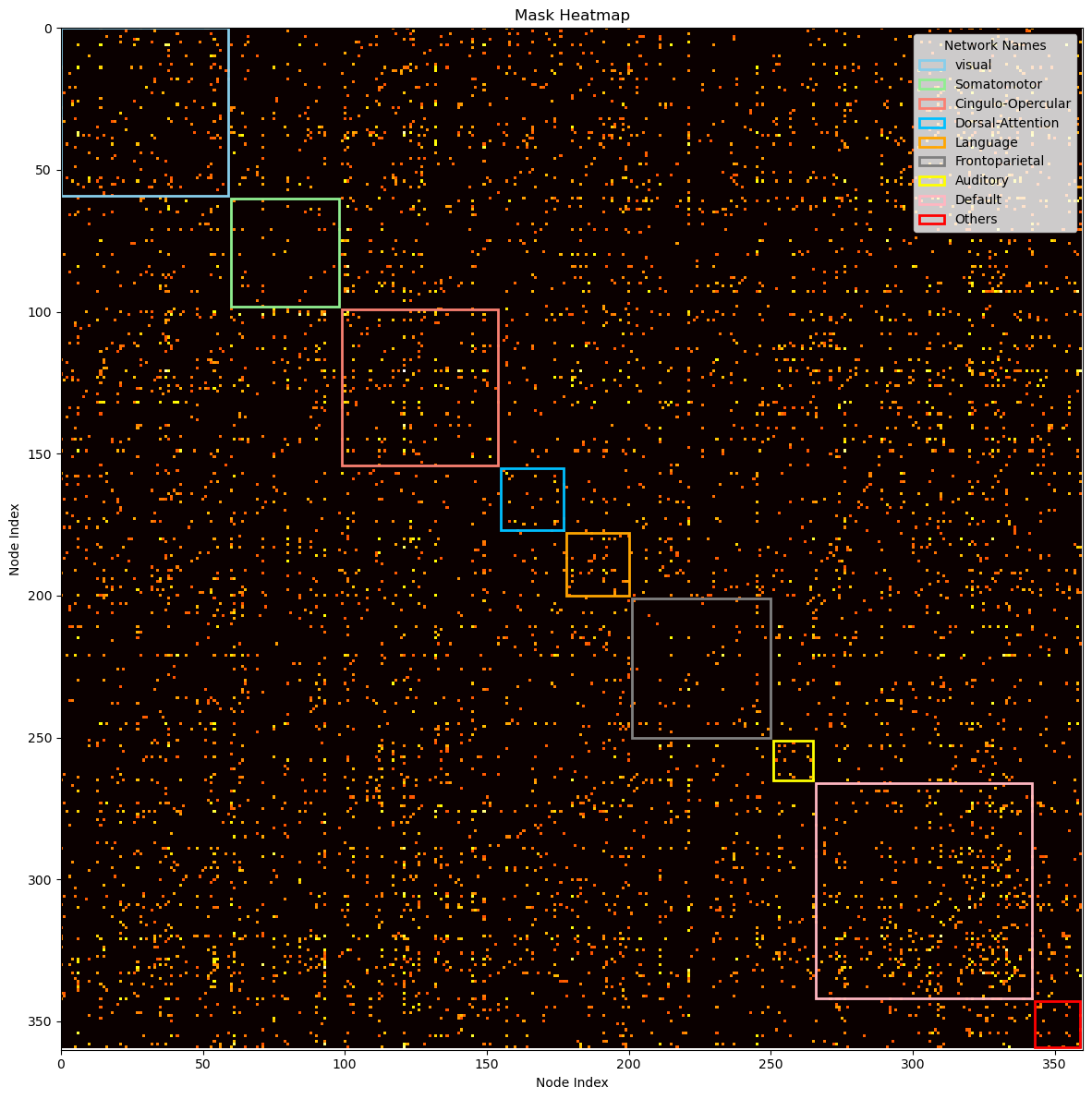}
  \label{fig:mask_class}
}
\subfloat[]{
  \includegraphics[width=0.4\textwidth]{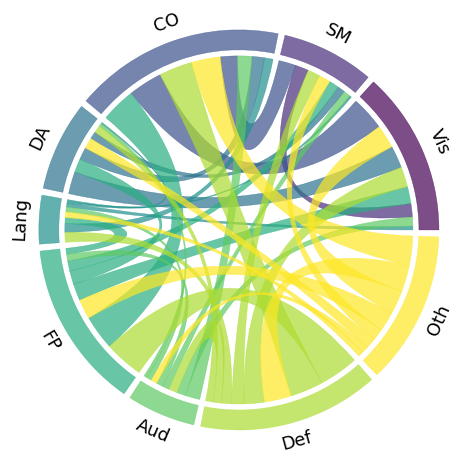}
  \label{fig:chord_diag_reg}
}
\caption{The model interpretability through learned masks with 0.52 as threshold: (a) mask derived from the simultaneous prediction task for CCC and FCC; (b) Chord diagram from the simultaneous CCC and FCC prediction task, showing inter-network connections among brain functional networks, excluding intra-network links;(c) mask generated for the classification task across distinct TCC levels;(d) Chord diagram from the classification task across distinct TCC levels, showing inter-network connections among brain functional networks, excluding intra-network links. Vis-Visual, SM-Somatomotor, CO-Cingulo-Opercular, DA-Dorsal-Attention, Lang-Language, FP-Frontoparietal
, Aud-Auditory, Def-Default, Oth-Others.}
\label{fig:mask}
\end{figure*}
\subsubsection{Brain Region Identification}
\begin{figure*}[htp!]
\centering
\subfloat[]{
  \includegraphics[width=0.48\textwidth]{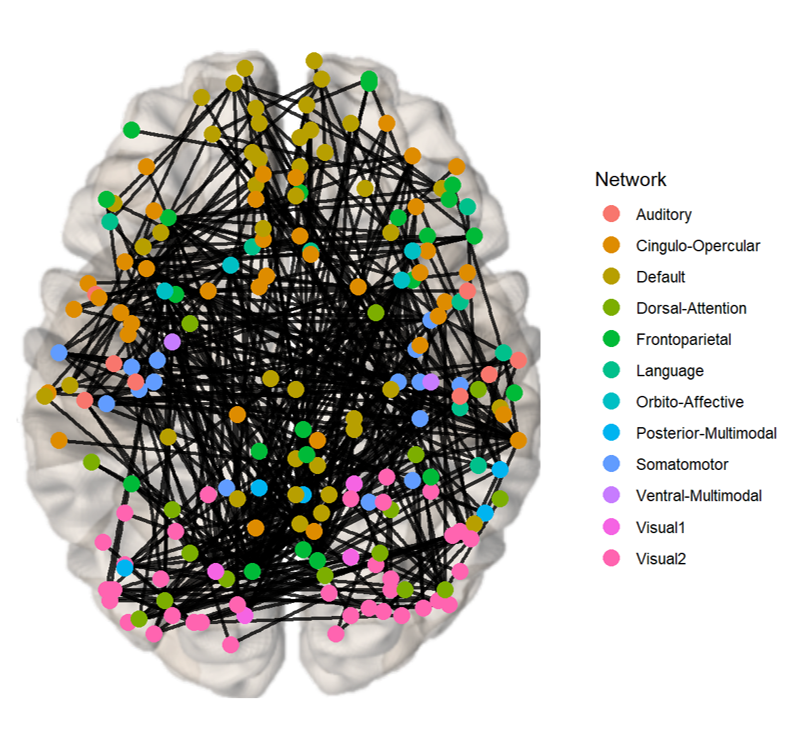}
  \label{fig:brainvis_reg}
}
\subfloat[]{
  \includegraphics[width=0.48\textwidth]{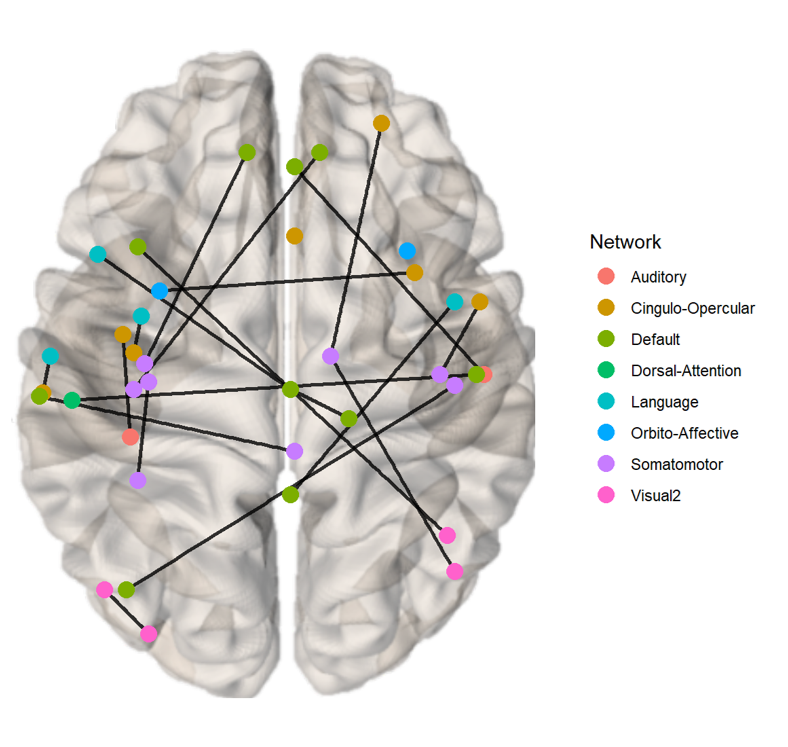}
  \label{fig:brainvis_class}
}
\caption{The visualization of Identified Brain Connectivities: Enhanced clarity is achieved by setting the visualization threshold to 0.53. (a) Connectivity patterns identified via the mask generated from the prediction task for CCC and FCC scores. (b) Connectivity patterns identified via the mask generated from the classification task for groups with high and low TCC scores.}
\label{fig:brainvis}
\end{figure*}

Drawing on existing knowledge, the connectivity network can be segmented into several brain functional networks: Visual, Somatomotor, Cingulo-Opercular, Dorsal-Attention, Language, Frontoparietal, Auditory, Default, and additional networks such as Posterior-Multimodal, Ventral-Multimodal, and Orbito-Affective. As illustrated in Fig.\ref{fig:mask_reg}, the majority of these brain functional networks participate in the cognition prediction task. However, it is noteworthy that the Auditory network and language network demonstrate a marked reduction in sparsity compared to others. Moreover, the network patterns observed in our findings exhibit slight deviations from the predefined functional networks. This is likely a result of incorporating both SC and FC while excluding subcortical regions, which extends the network identification beyond solely functional properties. Nevertheless, discernible network patterns remain evident in our analysis.
The brain connectivity patterns identified in this study are illustrated in Fig.\ref{fig:brainvis}. This visualization provides a comprehensive overview of the neural connections discovered through our analysis, offering insights into the complex network dynamics within the brain.

Furthermore, Fig.\ref{fig:mask} reveals differences in the sparsity of masks and the chord diagrams, which illustrate the interactions between brain functional networks for classification and regression tasks, even with a consistent threshold of 0.52. Classification models, aiming to distinguish discrete groups, prioritize a selective set of discriminative connectivities, leading to sparser visual representations. In contrast, regression tasks, focusing on continuous CCC and FCC scores, incorporate a broader range of connectivities for nuanced variation capture, resulting in denser representations even with the $L_1$ and $L_2$ sparsity terms.

By applying Grad-RAM to analyze the prediction experiments, it is observed from Fig.\ref{fig:gradcam_reg} that the average cortical thickness across all ROIs and the INT emerge as the top two anatomical statistics (AS) for predicting CCC and FCC scores. Meanwhile, other AS exhibit comparable Grad-RAM scores. Nonetheless, the regression task does not capture the differences among groups, where each AS may play a unique role depending on the group. Consequently, we further examine the Grad-CAM scores in classifying subjects into different TCC levels, highlighting how AS influences both the extremely high and borderline TCC group. The results presented in Fig.\ref{fig:gradcam_class} reveal that the number of vertices, gray matter volume, and structure-function coupling emerge as the most distinctive features.

To assess the impact of atlas selection on our results, we performed additional analyses employing the Schaefer 200 atlas \citep{schaefer2018local} for brain parcellation. The findings, detailed in the \ref{app:Schaefer}, demonstrate a high degree of consistency across different atlases. We identified the same top important features with only minor discrepancies noted, confirming the robustness of our methodology. However, the distinct patterns observed from the learned mask were less pronounced when using the Schaefer atlas compared to those derived from the Glasser atlas. This variation can be attributed to the Schaefer atlas’s adaptable clustering options and its suboptimal configuration for integrating multimodal imaging data. These results highlight the benefits of utilizing a multimodal-specific atlas, such as the Glasser atlas to enhance clarity in network delineation.

\subsection{Discussion}
\label{Discussion}
\subsubsection{Interactions Between Cognitive Networks and Intelligence}
Our findings from Fig.\ref{fig:mask_reg} indicate a relatively lower density in language and auditory brain functional networks in the context of predicting crystal and fluid intelligence, suggesting these networks are related yet not closely tied to the core domains of general intelligence. Given the participants' healthy status, the findings likely represent group-level variations rather than outcomes related to developmental challenges in language and academics found in individuals with hearing impairments \citep{heinrichs2022auditory}. Supporting evidence from the study \citep{woolgar2018multiple} on the multiple-demand (MD) system and frontoparietal brain regions further clarifies this distinction. While the MD system's association with fluid intelligence highlights the importance of domain-general regions, the differential effects of lesions in these areas indicate a specific link between the MD system and fluid intelligence, rather than language processing alone. The distinction between nonverbal intelligence, separate from academic intelligence, and speech intelligence, linked to verbal reasoning, underscores cognitive diversity. Confirmatory factor analyses reveal auditory nonverbal intelligence as a distinct domain, suggesting the inclusion of a nonverbal auditory dimension in intelligence models could deepen our understanding of cognitive functions. This is supported by research showing nonverbal and speech abilities contribute uniquely to cognitive profiles, highlighting the importance of auditory processes in intelligence frameworks.

Moreover, results illustrated in Fig.\ref{fig:mask_class} demonstrate clear differences in Grad-CAM scores between high and low total in groups across the default mode network and cingulo-opercular network, linking cognitive abilities to distinct connectivity patterns in these networks. The association of higher-order cognitive abilities with the efficiency of the cingulo-opercular network underscores its critical role in cognitive performance. This connection is further highlighted by the impact of psychotic-like experiences (PLEs) on network efficiency and the mediation of cognitive ability by cingulo-opercular network efficiency \citep{sheffield2016cingulo}, emphasizing the cingulo-opercular network's centrality in cognitive functioning. Our analysis aligns with previous research  \citep{hearne2016functional, song2009default, pamplona2015analyzing, santarnecchi2015intelligence} showing individual intelligence differences related to changes in resting state connectivity across networks engaged in self-referential mental activity (default mode network) and task-set maintenance (cingulo-opercular network), reinforcing the significance of connectivity variations in influencing cognitive outcomes. This relationship underscores the importance of network efficiency in cognitive health and suggests that even subtle differences in the connectivity within these networks can have substantial implications on cognitive performance.

\subsubsection{Limitations and Prospects}
Regarding the proposed method, we have delineated a framework for multimodal analysis of neuroimaging data, which has been a big challenge in integrating sMRI, fMRI and DTI. Although it demonstrates high efficiency in analyzing multiple modalities of neuroimaging, the methodology employed for achieving specific outcomes warrants further refinement. Firstly, the strategy utilized for the fusion of fMRI and DTI data at the nodal level, coupled with the assimilation of the AS into the latent layer via concatenation, is preliminary. The integration of more advanced fusion techniques, including those utilizing attention mechanisms \citep{xie2023multimodal, nagrani2021attention} and incorporating generative models \citep{jin2025graph, guan2025spatio, orlichenko2023angle}, has the potential to enhance the effectiveness of the proposed framework. Secondly, the procedure for initializing the mask, which is a variation on our previous work, utilizes a basic approach. Alternative methodologies, such as low-rank matrix factorization, could offer improvements with the incorporation of additional prior knowledge into the analysis. Furthermore, while our framework has proven to be effective across different atlas settings, other factors such as the variation in data processing techniques or the integration of newer modalities may still need further validation to confirm their impact on the model’s performance and interpretability. These areas present opportunities to refine and enhance the robustness of our approach, potentially improving its predictive accuracy and applicability in a broader neuroscientific context. Additionally, we concentrate on the analysis of the HCP-D dataset, which includes only healthy individuals. Therefore, the applicability of our findings to populations with cognitive deficits or neurological disorders remains uncertain. Expanding our analysis to encompass datasets featuring subjects across a spectrum of cognitive impairments would gain additional insights and stand as an interesting direction for further study.
\section{Conclusion}
\label{Conclusion}
In this research, we introduced an integrated multimodal neuroimaging framework utilizing MaskGNN to synergize heterogeneous imaging data including fMRI, DTI, and sMRI.
To our knowledge, this work is among a handful of studies to successfully integrate fMRI, sMRI, and DTI within a novel deep learning framework. This novel approach not only harmonizes disparate data into a cohesive analytical framework but also exploits the unique strengths of each imaging modality to unravel the complexities of brain connectivity, structure and function. Our methodology, rigorously validated on the HCP-D dataset, demonstrates the importance of combining FC, SC, and AS to significantly enhance predictive accuracy in cognitive function mapping. Furthermore, by employing interpretability techniques such as learned masks and Grad-RAM/Grad-CAM analyses, we identified crucial brain connections and anatomical markers pivotal for cognitive processing. These findings affirm the efficacy of our integrated approach and provide new perspectives on the interplay between the brain's network dynamics and cognitive functionalities.
In conclusion, our work introduces a novel framework for the integrated examination of multimodal imaging data and for delineating the intricate relationships between the brain's structural and functional networks and their influence on cognitive development.
%It also sets a fertile ground for future investigations to further exploit these complex interconnections.
\section*{Data and Code availability for replication}

The code is openly available at \url{https://github.com/GQ93/IBrainGNN_fMRI_DTI_sMRI}. Data cannot be open-sourced due to restrictions but can be provided upon special request.
\section*{Acknowledgments}
This work was supported in part by NIH under Grants R01 GM109068, R01 MH104680, R01 MH107354, P20 GM103472, R01 REB020407, R01 EB006841, U19AG055373, and in part NSF under Grant $\#$1539067.
\section*{Author contributions statement}
Gang Qu was responsible for the principal data analysis, coding execution, conducting experiments, and the composition and critique of the manuscript. Aiying Zhang engaged in conceptualizing the project, data processing, and provided critical reviews of the manuscript. Ziyu Zhou and Vince D. Calhoun contributed to the evaluation of the model's design and offered valuable recommendations during the manuscript review process. Yu-Ping Wang was responsible for conceptualizing the project, securing funding, and reviewing the manuscript.
\section*{Declaration of generative AI and AI-assisted technologies in the writing process}
During the preparation of this work, the authors used OpenAI ChatGPT 4 to improve the readability and language of the manuscript. This technology was used with strict human oversight, with the authors thoroughly reviewing and revising the output to ensure accuracy and integrity. The authors confirm that AI tools were not used to generate scientific content and are not listed as authors or co-authors. After using this tool/service, the authors reviewed and edited the content as needed and take full responsibility for the content of the published article.
\section*{Financial Disclosures}

All authors declare that they have no conflicts of interest.

\appendix
\section{Structure-function Coupling}
\label{app:sf}
\setcounter{figure}{0}
To enrich the analytical robustness of the AS features, we augment the feature vector on each ROI  with structure-function coupling, employing the Spearman rank-order correlation coefficient \citep{baum2020development}  to quantitatively assess the relationship between FC and SC. Given a set of FC and SC vectors for a ROI, the Spearman rank-order correlation coefficient $\rho$, is shown in Eq.\ref{rho}.

\begin{equation}
\rho = 1 - \frac{6 \sum d_i^2}{n(n^2 - 1)},
\label{rho}
\end{equation}
where $d_i$ denotes the difference between the ranks of corresponding FC and SC values, and $n$ is the total number of observations.

The calculation sequence for the Spearman rank-order correlation coefficient, $\rho_i$, followed by the augmentation of the AS feature sets is shown below:

\begin{enumerate}
\item Assign ranks to both FC and SC values.
\item Determine the rank difference, $d_i$, for each corresponding FC and SC pair.
\item Compute the square of each rank difference, yielding $d_i^2$.
\item Aggregate these squared differences to produce $\sum d_i^2$.
\end{enumerate}

\section{Revalidation Results Using the Schaefer Atlas}
\label{app:Schaefer}
\begin{figure}[ht!]
\centering
\subfloat[]{
  \includegraphics[width=0.4\textwidth]{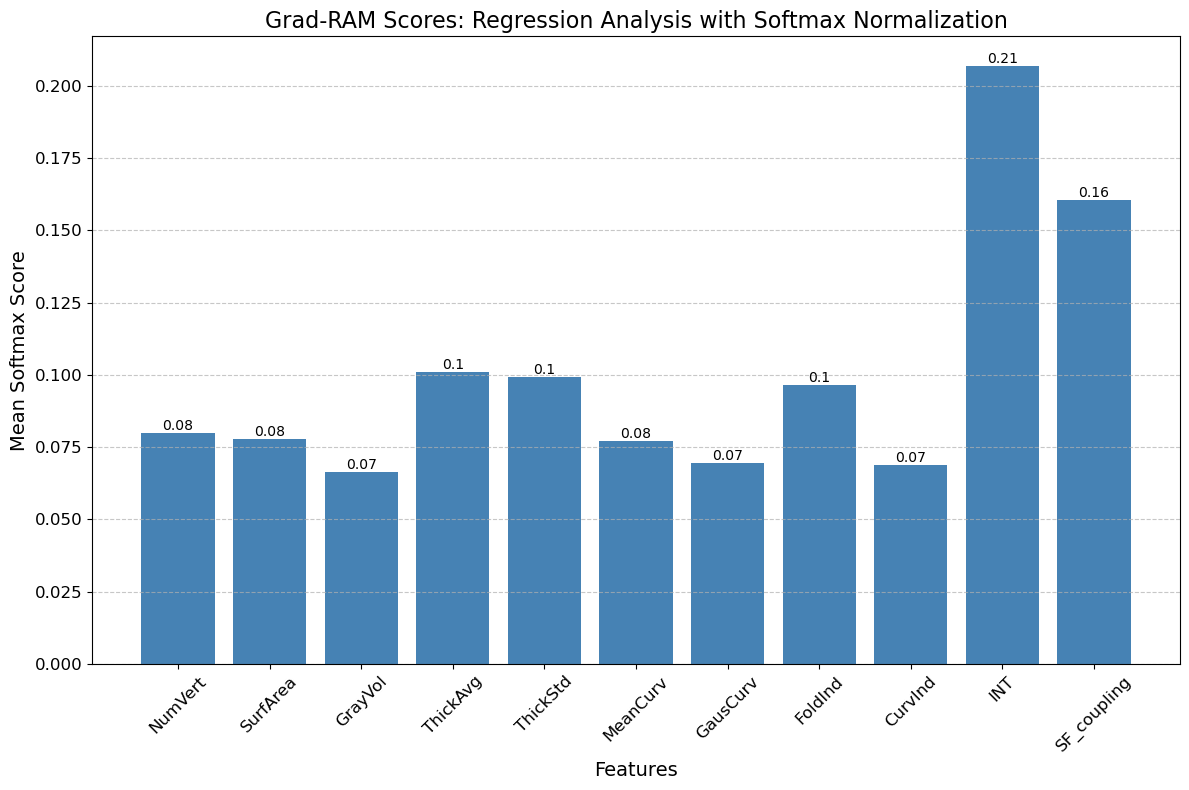}
  \label{fig:gradcam_reg_Schaefer}
}
\hspace{0.1cm}
\subfloat[]{
  \includegraphics[width=0.4\textwidth]{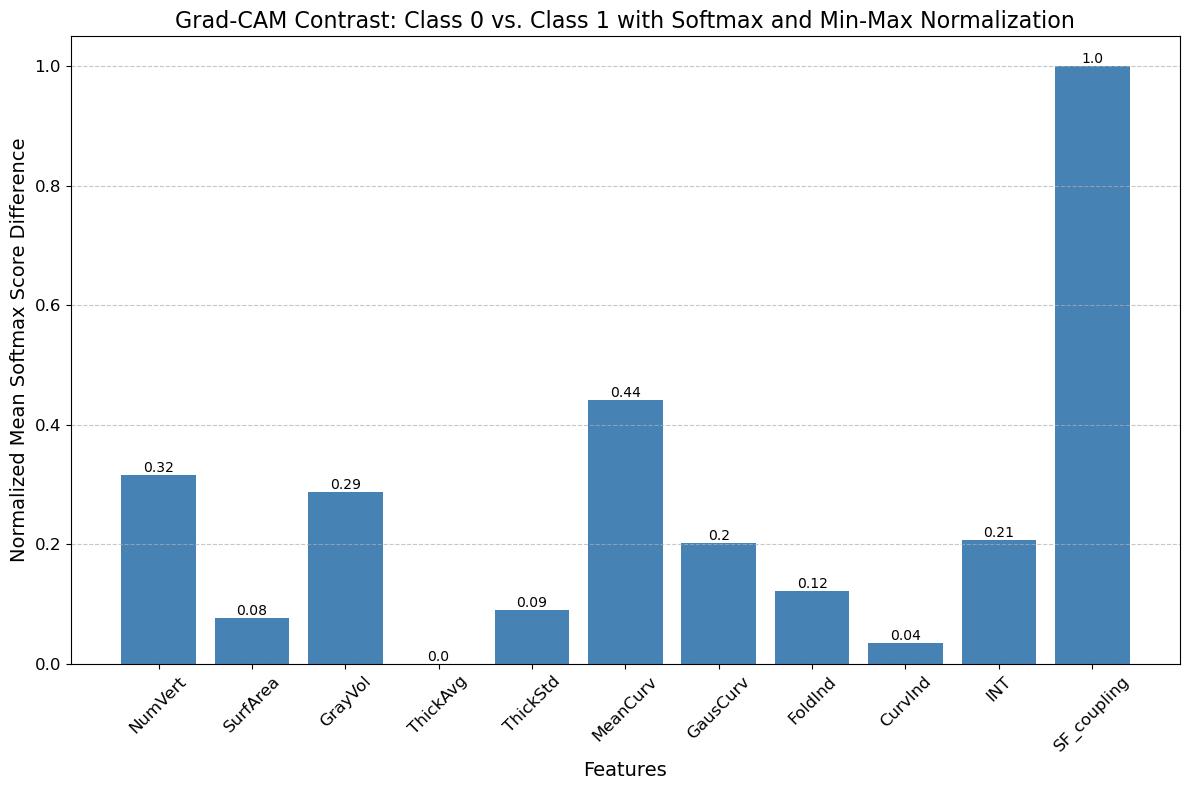}
  \label{fig:gradcam_class_Schaefer}
}
\caption{The use of Grad-CAM and Grad-RAM scores for model explainability, utilizing Schaefer atlas for brain parcellation: (a) Grad-RAM scores for simultaneous prediction of CCC and FCC; (b) Discrimination of groups using Grad-CAM scores across distinct TCC levels.}
\label{fig:gradscore_Schaefer}
\end{figure}

\begin{figure}[h!]
\centering
\subfloat[]{
  \includegraphics[width=0.2\textwidth]{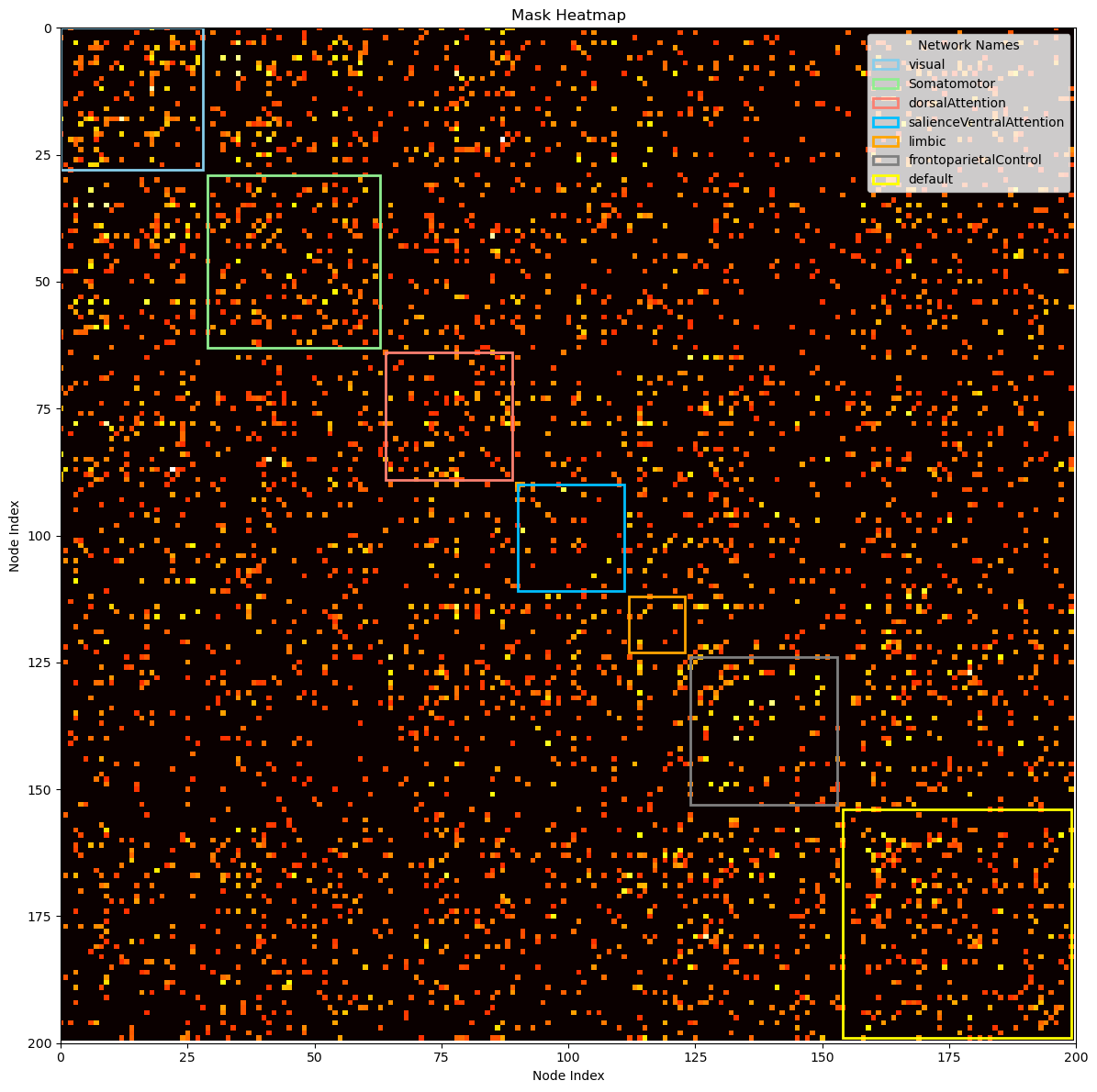}
  \label{fig:mask_reg_Schaefer}
}
\hspace{0pt}
\subfloat[]{
  \includegraphics[width=0.2\textwidth]{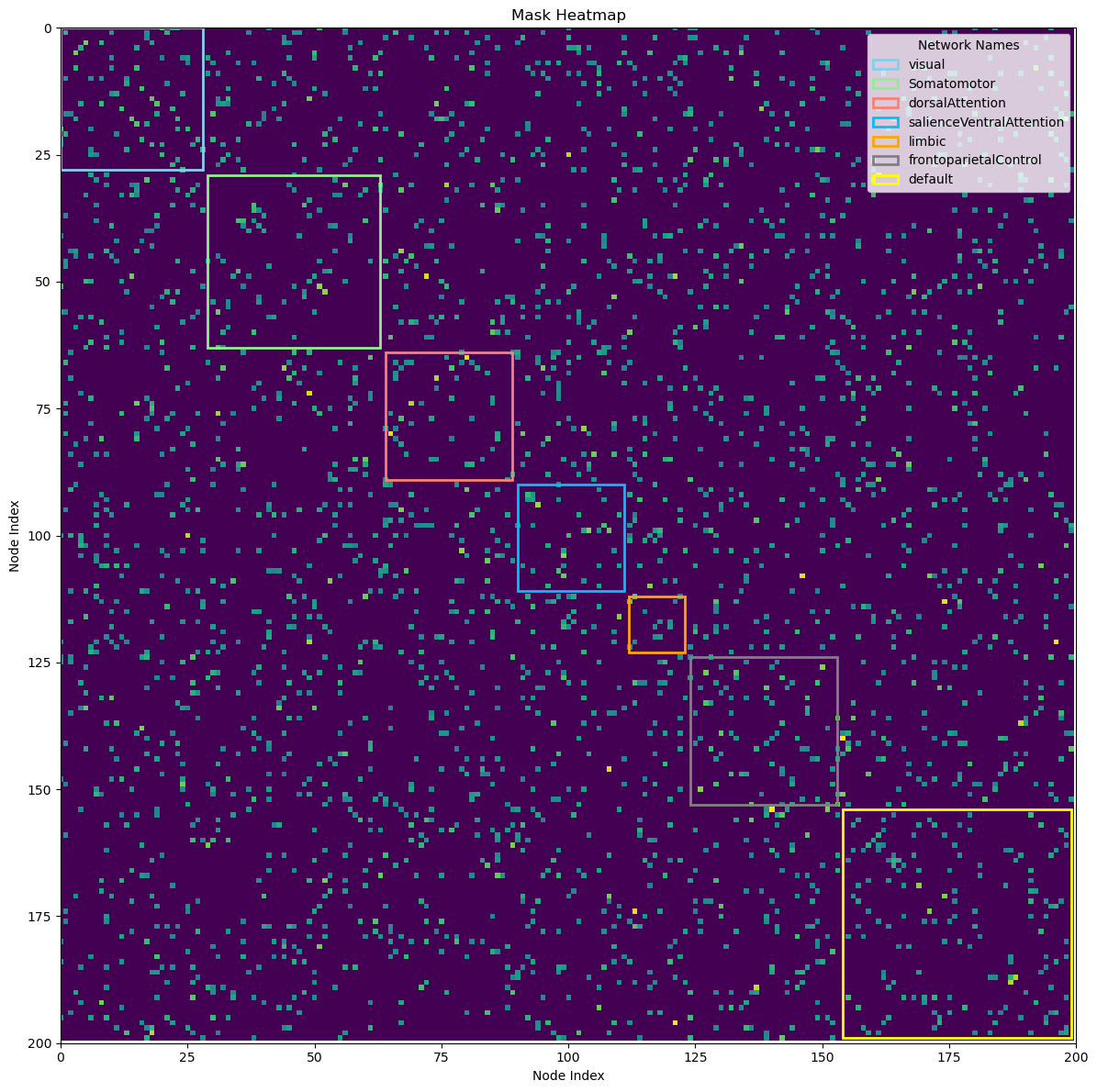}
  \label{fig:mask_class_Schaefer}
}
\caption{The model interpretability through learned masks with 0.52 as threshold, utilizing Schaefer atlas for brain parcellation: (a) mask derived from the simultaneous prediction task for CCC and FCC;(b) mask generated for the classification task across distinct TCC levels.}
\label{fig:mask_Schaefer}
\end{figure}

We revalidated our framework using the Schaefer 200 atlas for brain parcellation. As shown in Figure \ref{fig:gradcam_reg_Schaefer}, the results largely corroborate our initial findings, with INT remaining as the most crucial feature for predicting CCC and FCC scores. Unlike previous results where average cortical thickness was significant, structure-function coupling emerged as a key feature.
However, metrics for cortical thickness (average and standard deviation) still rank highly when combined, as seen in their Grad-CAM scores. For classification tasks distinguishing TCC levels, Figure \ref{fig:gradcam_class_Schaefer} shows minimal variation with consistent top features. The masks derived using the Schaefer atlas, depicted in Figures \ref{fig:mask_reg_Schaefer} and \ref{fig:mask_class_Schaefer}, display less distinct patterns compared to the Glasser atlas. This could be due to the Schaefer atlas's flexible clustering and non-optimization for multimodal imaging, potentially affecting the clarity of network delineation and masking patterns.

% \section{Hyperparameter Sensitivity}
% \label{app:Hyperparameter}
\clearpage
\bibliographystyle{model2-names.bst}\biboptions{authoryear}
\bibliography{refs}

\end{document}